\documentclass[onecolumn]{aa}   
\usepackage{graphicx}   
\begin{document}   
   \title{Photometric structure of the peculiar galaxy ESO~235-G58}   
   
   \author{E. Iodice,   
          \inst{1}   
          \and   
           M. Arnaboldi   
          \inst{2,1}   
          \and    
           L. S. Sparke   
          \inst{3} 
          \and    
           R. Buta   
          \inst{4}         
          \and  
           K.C. Freeman  
          \inst{5}  
          \and     
           M. Capaccioli   
          \inst{1,6}        
                    }   
   
   \offprints{E. Iodice, \email{iodice@na.astro.it}}   
   
   \institute{INAF-Osservatorio Astronomico di Capodimonte (OAC),    
              via Moiariello 16, I-80131 Napoli   
         \and  
	      INAF-Osservatorio Astronomico di Torino,    
              Strada Osservatorio 20, I-10025 Pino Torinese  
         \and      
             Dept. of Astronomy,   University of Wisconsin, 
             475 N. Charter St., Madison, WI   
         \and     
          University of Alabama, Department of Physics \& Astronomy,   
           Box 870324, Tuscaloosa, Alabama 35487  
         \and  
          RSAA, Mt. Stromlo Observatory, Canberra, Cotter Road,\\  
           Weston ACT 2611, Australia  
         \and     
	   Dept. of Physical Sciences, University ``Federico II'',  
              Naples, Italy  
             }   
   
   \date{Received; accepted }

\abstract{We present the near-infrared and optical properties of the 
peculiar galaxy ESO~235-G58, which resembles a late-type ringed barred
spiral seen close to face-on. However, the apparent bar of ESO~235-G58
is in reality an edge-on disk galaxy of relatively low luminosity.  We
have analyzed the light and color distributions of ESO~235-G58 in the
NIR and optical bands and compared them with the typical properties
observed for other morphological galaxy types, including polar ring
galaxies. Similar properties are observed for ESO~235-G58, polar ring
galaxies, and spiral galaxies, which leads us to conclude that this
peculiar system is a {\it polar-ring-related} galaxy, characterized by
a low inclined ring/disk structure, as pointed out by Buta \& Crocker
in an earlier study, rather than a barred galaxy.
   
}   
    
\titlerunning{} \authorrunning{Iodice et al.}   
                                                
   \maketitle   
%
   
\section{Introduction}\label{intro}   
   
The southern galaxy ESO~235-G58 is classified in RC3 (de Vaucouleurs
et al. 1991) as an SB(rs)d spiral based on its appearance on the
ESO/SRC-J Southern Sky Survey. It is characterized by a quite
elongated bar-like structure, for which the total extension is $\sim
40''$ and the position angle (P.A.) of its major axis is
$106^\circ$. This apparent bar is encircled by a weak inner
pseudo-ring and two faint outer arms, which reach a distance of about
$1'$ from the center (Fig.\ref{eso235B}). Using $H_0=70$ km s$^{-1}$
Mpc$^{-1}$, the galaxy has a distance of about $67$ Mpc. In this early
classification the galaxy disk is seen nearly face-on, with a
pseudo-ring at about 1' from the center, encircling an inner bar.
   
Closer inspection of ESO~235-G58 (Buta 1995), in the {\it Catalogue of
Southern Ringed Galaxies}, led to the suspicion that the apparent
central bar is in reality an edge-on disk. With new optical data, Buta
and Crocker (1993) showed that ESO~235-G58 is probably not a ringed
barred spiral of late Hubble type, because the central elongated
structure shows a nearly linear dust lane which splits the nuclear
region in a way which is typical of edge-on disks (see
Fig.~\ref{eso235B}). This implies that the equatorial plane of the
central structure is at a different angle to the line of sight than
the outer ring.  The analysis of these optical data (Buta and Crocker
1993) suggested that ESO~235-G58 is composed of a low luminosity
edge-on disk, with an equatorial dust-lane, and an outer ring/spiral
at very large radii, seen at a lower inclination. Therefore,
ESO~235-G58 may be a good candidate for Polar-Ring-related
objects. Polar Ring galaxies (PRGs) are usually recognized because the
two components (the host galaxy and ring) are nearly edge-on (Whitmore
et al. 1990). It is more difficult to recognize cases where the inner
component is edge-on, while the ring is more face-on: cases like that
must exist and ESO235-G58 may be such kind of an object. 
 
The main goal of the present work is to provide accurate optical and 
near-infrared (NIR) photometry of ESO~235-G58, and to compare the main 
properties of this galaxy with those typical of other morphological 
galaxy types, including PRGs.  NIR photometry is necessary to minimize 
the dust absorption which strongly affects the starlight distribution 
in the central galaxy. In addition, the study of optical and NIR 
integrated colors will provide information about the age and 
metallicity of the dominant stellar population in the different 
components of the system.  In this paper we present new NIR 
observations, obtained for ESO~235-G58 in J, H and Kn bands, and have 
applied the same procedures adopted to study PRGs by Iodice et 
al. (2002a, 2002b, 2002c). As shown by Iodice and collaborators, this 
kind of analysis allows us to obtain quantitative morphology of the 
main components in ESO~235-G58, in order to rigorously classify this 
object, and gain insight on its star formation history. 
   
Observations and data reduction are presented in Sec.\ref{obs}; the 
morphology, light and color distribution of the two components (host 
galaxy and ring) are discussed in Sec.\ref{eso235_morph} and 
Sec.\ref{eso235_phot}. In Sec.\ref{eso235_dust} we describe the dust 
properties and how they compare with the typical properties in other 
galaxies and the Milky Way. In Sec.\ref{eso235_col} the integrated 
colors derived for the central galaxy and ring are presented, and in 
Sec.\ref{age_eso235} we give an age estimate of the two components. 
The two-dimensional model of the central galaxy light distribution is 
discussed in Sec.\ref{eso235_g}, and in Sec.\ref{eso235_r} we discuss 
the properties of the ring light distribution. A summary of the main 
results and conclusions is presented in Sec.\ref{eso235_sum}. 
   

\section{Observations}\label{obs}   
   
ESO~235-G58 is a member of a selected sample of peculiar galaxies 
observed in the NIR J, H and Kn\footnote{The central wavelength for Kn 
filter is 2.165 $\mu m$.} bands with the CASPIR infrared camera 
(McGregor 1994), attached to the Mt. Stromlo and Siding Spring 
Observatory 2.3 m telescope. The pixel scale is 
$0.5''$ pixel$^{-1}$ and the field of view is $2.0' \times 2.0'$. A 
detailed description of the data reduction is given by Iodice et 
al. (2002b).  The observing log for these data is summarized in 
Table~\ref{obslog}. 
    
The optical data were obtained by D. A. Crocker and L. V. Jones in 
1992 with a TEK 1024 CCD array (with a pixel scale of $0.43''$ pixel$^{-1}$) 
attached to the 1.5 m telescope of Cerro Tololo Inter-American 
Observatory. The images were acquired in the Johnson BV and Cousin I 
filters, with total exposures of $1500 \mbox{s}$ in B, $600 \mbox{s}$ 
in V and $300 \mbox{s}$ in I.  The reduction and analysis of these 
data were published by Buta 
\& Crocker (1993). The B band and H band images are shown in   
Fig.\ref{eso235B} and Fig.\ref{eso235h} respectively.   
   
\begin{table}   
\centering    
\caption[]{NIR log of observations for ESO~235-G58.}   
\label{obslog}   
\begin{tabular}{cccc}   
\hline\hline   
Filter & Tot. int. & FWHM & Date \\   
       &  (s)      &(arcsec)&    \\   
\hline   
J  & 1200& 1.6& 19/08/1995\\   
H  & 1200& 1.5& 19/08/1995\\   
H  &  600& 1.6& 20/08/1995\\   
Kn  & 2400& 1.5& 18/08/1995\\   
Kn  &  600& 1.4& 20/08/1995\\   
\hline   
\end{tabular}   
\end{table}

   
\section{Morphology in the NIR and optical bands}\label{eso235_morph}   
   
\begin{figure*}
\includegraphics[width=13cm]{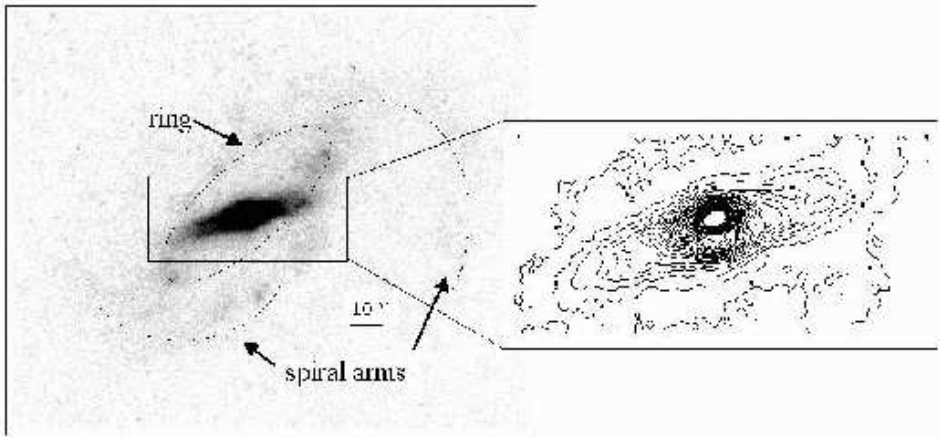}   
\caption{\label{eso235B} Left panel - B-band image, by Buta and Crocker (1993),   
 of the peculiar galaxy ESO~235-G58. Right panel - Contour plot of the   
 inner regions of ESO~235-G58, strongly perturbed by a dust lane which   
 splits the galaxy along the major axis.  North is at the top and   
 East is to the left. Units are intensity.}   
\end{figure*}

\begin{figure*}
\resizebox{\hsize}{!}{ 
\includegraphics[width=3cm]{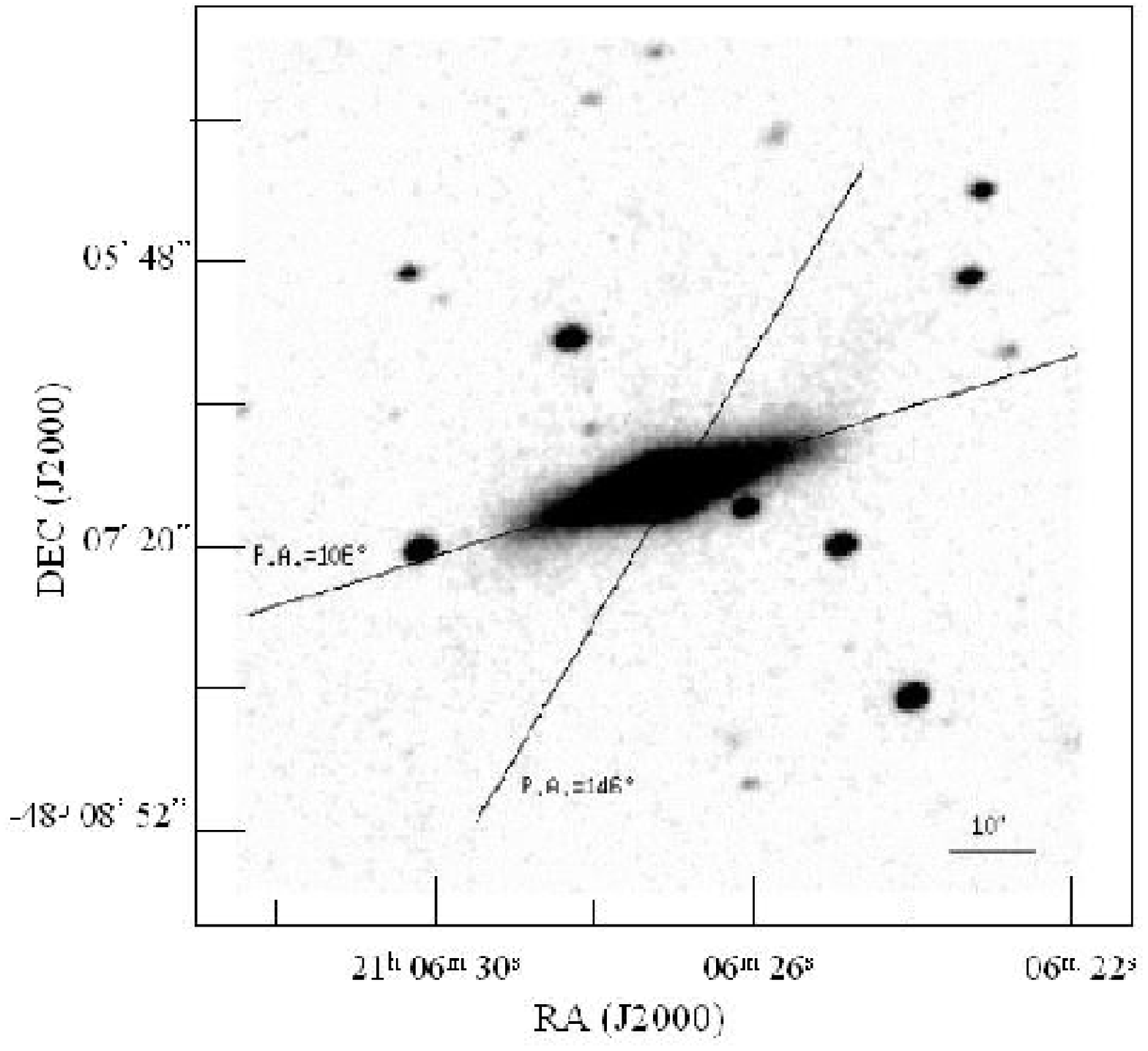}  
\includegraphics[width=2.5cm]{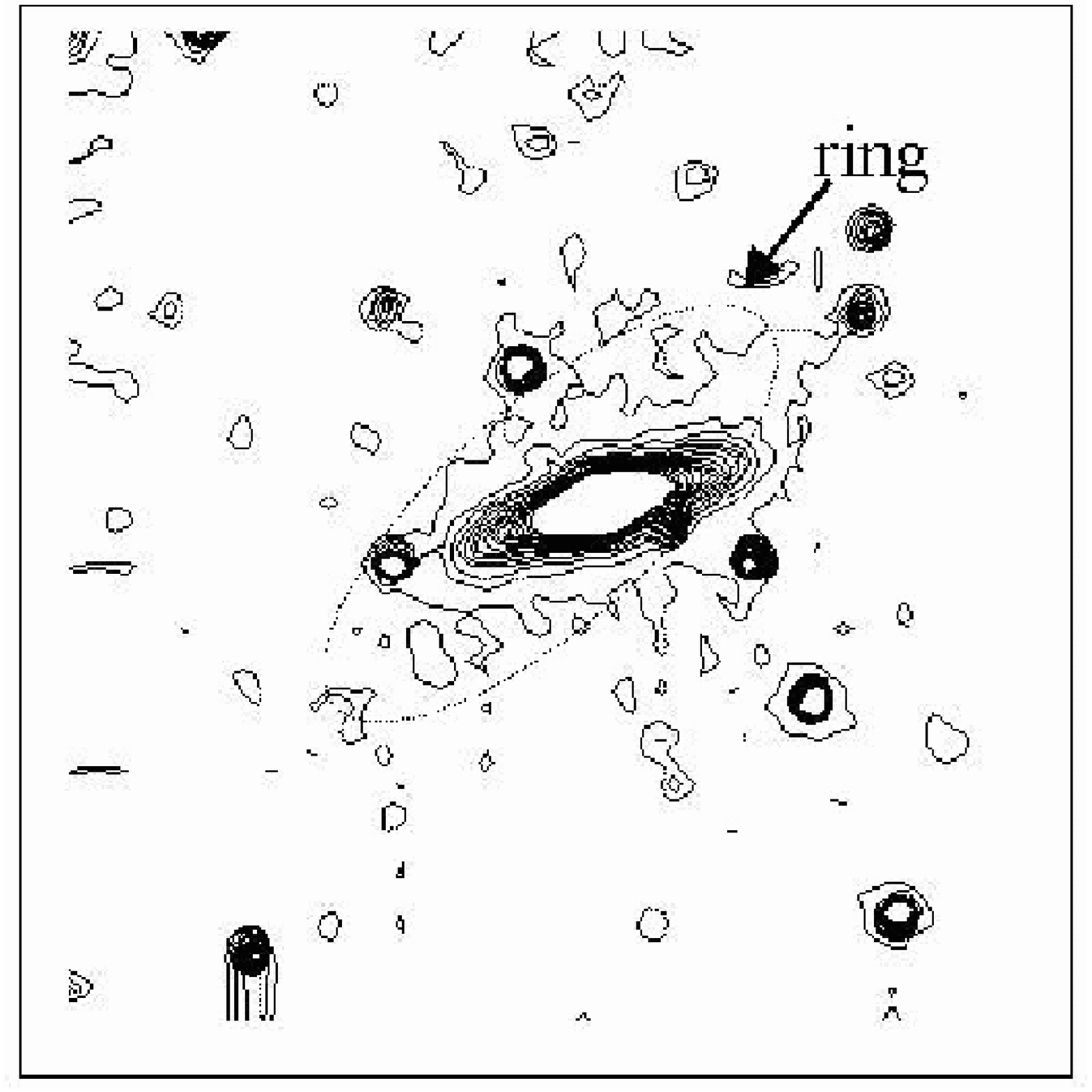} 
}   
\resizebox{\hsize}{!}{ \includegraphics[width=3cm]{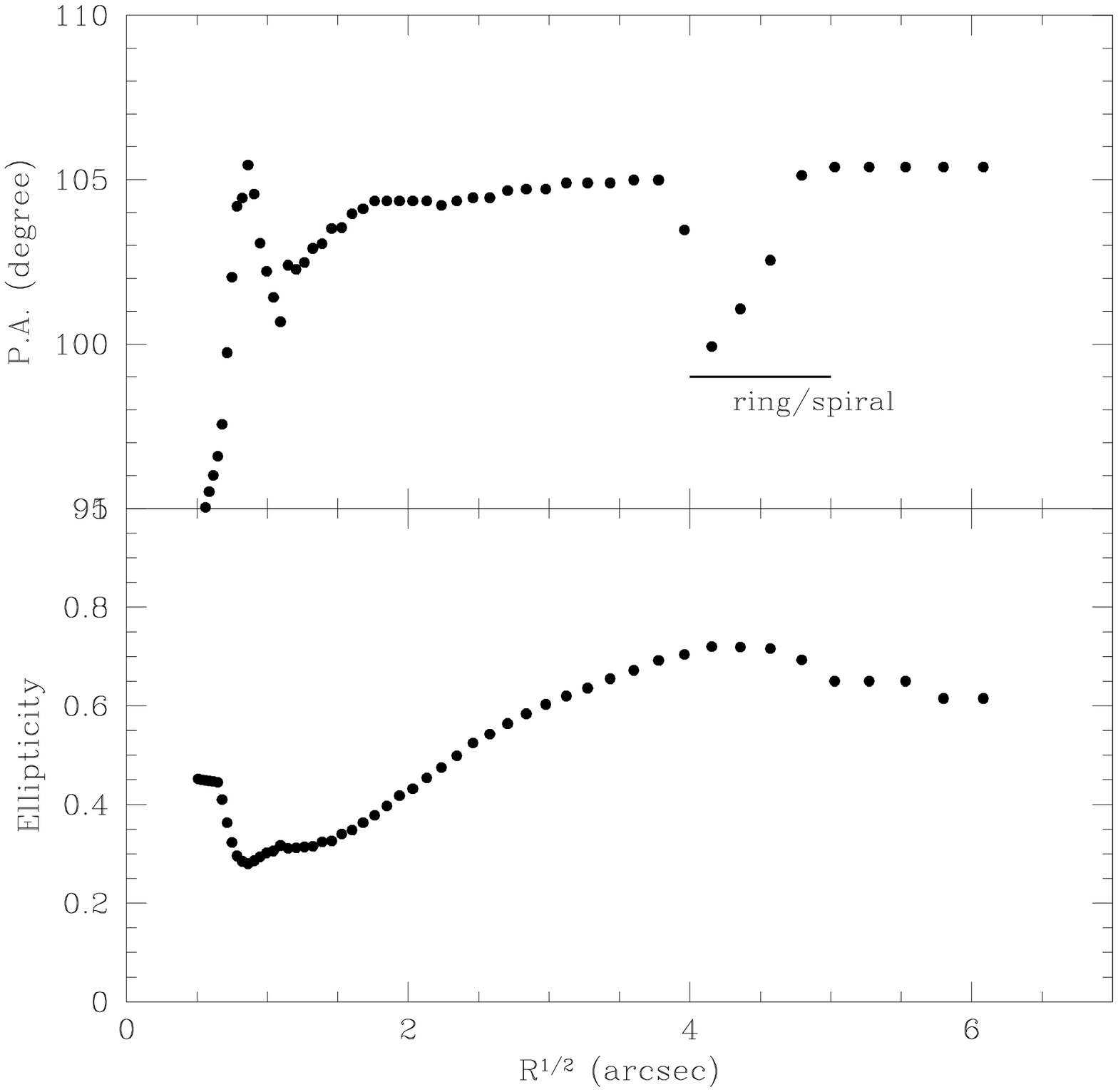} 
\includegraphics[width=3cm]{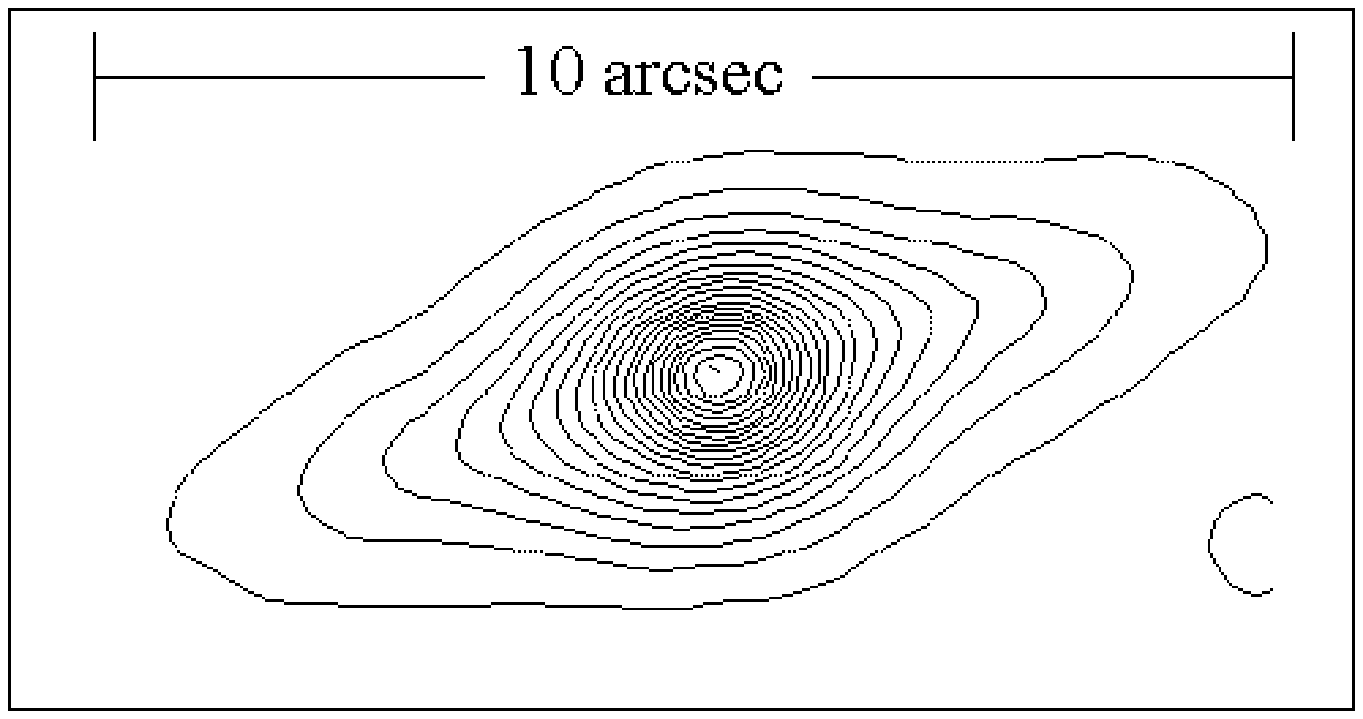}  
}
\caption{\label{eso235h} Top-left panel: H-band image of 
ESO~235-G58; top-right panel: contour plot, with linear spacing, from
the intensity level corresponding to 16.25 mag/arcsec$^2$ to
background level. The lines at $P.A.=106^o$ and $P.A.=146^o$ show
apparent major axes of the central galaxy and the outer ring
respectively. The image size is $2' \times 2'$. North is at the top
and East is to the left.  In the bottom-left panel are shown the
ellipticity and P.A. profiles along the major axis of the galaxy.  In
the bottom-right panel is shown the contour plot, with a linear
spacing, of the inner regions ($R < 5''$) of the central galaxy, 
the last isophote corresponds to 16.45 mag/arcsec$^2$.}
\end{figure*}

The H band image of ESO~235-G58 is shown in Fig.~\ref{eso235h}. The 
appearance of the galaxy in this band is remarkably different from 
what is observed in the B band (Fig.\ref{eso235B}). The NIR morphology 
of ESO~235-G58 resembles that of an early-type, edge-on disk galaxy, 
similar to an Sa, while the ring and spiral arms, observed in the 
optical images, are not detected. The only NIR flux associated with 
this component is an asymmetric elongation, in the NW direction, of 
the central Sa disk (Fig.\ref{eso235h}). This structure may be debris 
in the apparent ring feature, which, on the other hand, is evident in 
the B (Fig.\ref{eso235B}) and V bands (see Buta \& Crocker 1993). This 
indicates that the ring structure is much bluer than the central 
object. 
   
The absence of squared-off ends, typical of bars in early-type spirals
(Athanassoula et al. 1990; Jungwiert et al. 1997; Erwin \& Sparke
2003), in the H band isophotes (see Fig.\ref{eso235h}, right panels),
make this object more similar to an edge-on disk galaxy than to a
bar. This is further confirmed by the shape of the ellipticity and
position angle profiles (shown in Fig.\ref{eso235h}, bottom-left
panel), derived by fitting ellipses\footnote{We have used the IRAF
task ELLIPSE to derive the ellipticity and P.A. profiles.} to
isophotes of the H band image: the ellipticity profile suggests the
existence of a round central component and an outer, more extended,
flattened structure. The position angle is almost constant except for
a variation of about four degrees in the range $16'' \le R \le 20''$.
The ellipticity and P.A. profiles are quite irregular in the innermost
regions ($R \le 3''$): the first maximum value in the ellipticity
corresponds to a minimum in the P.A., at about $1''$ from the center.
These variations could be caused by uncertainties in the fitting
algorithm, induced by the few adopted points, as it is also suggested
by the very regular aspect of the the inner isophotes, relative to
this region (see Fig.\ref{eso235h}, bottom-right panel). The
ellipticity and P.A. profiles for ESO~235-G58 are quite different from
those typically observed for barred galaxies, where the bar isophotes
are very nearly circular in the innermost parts (it varies from 0.7 to
0.9) and gradually become more elongated with increasing radius
(Athanassoula et al. 1990; Jungwiert et al. 1997); at the same time, a
twisting is observed in the P.A. profile (Wozniak et al. 1995; Erwin
\& Sparke 2003). In the case of ESO~235-G58, the twisting observed in
the P.A. profile, in the range $16'' \le R \le 20''$
(Fig.\ref{eso235h}), is due to the projection of the ring light on to
the central galaxy. The absence of a local maximum in the ellipticity
profile in this range suggest that there is no additional component,
like a bar, present. On the other hand, the ellipticity and
P.A. profiles for ESO~235-G58 are very similar to the typical
ellipticity and P.A. profiles observed for edge-on non-barred disk
galaxies (S0s and spirals), where an almost round bulge and a
prominent disk are present (Michard 1984; Scorza \& Bender 1995).\\
   
\begin{figure*}   
\includegraphics[width=6cm]{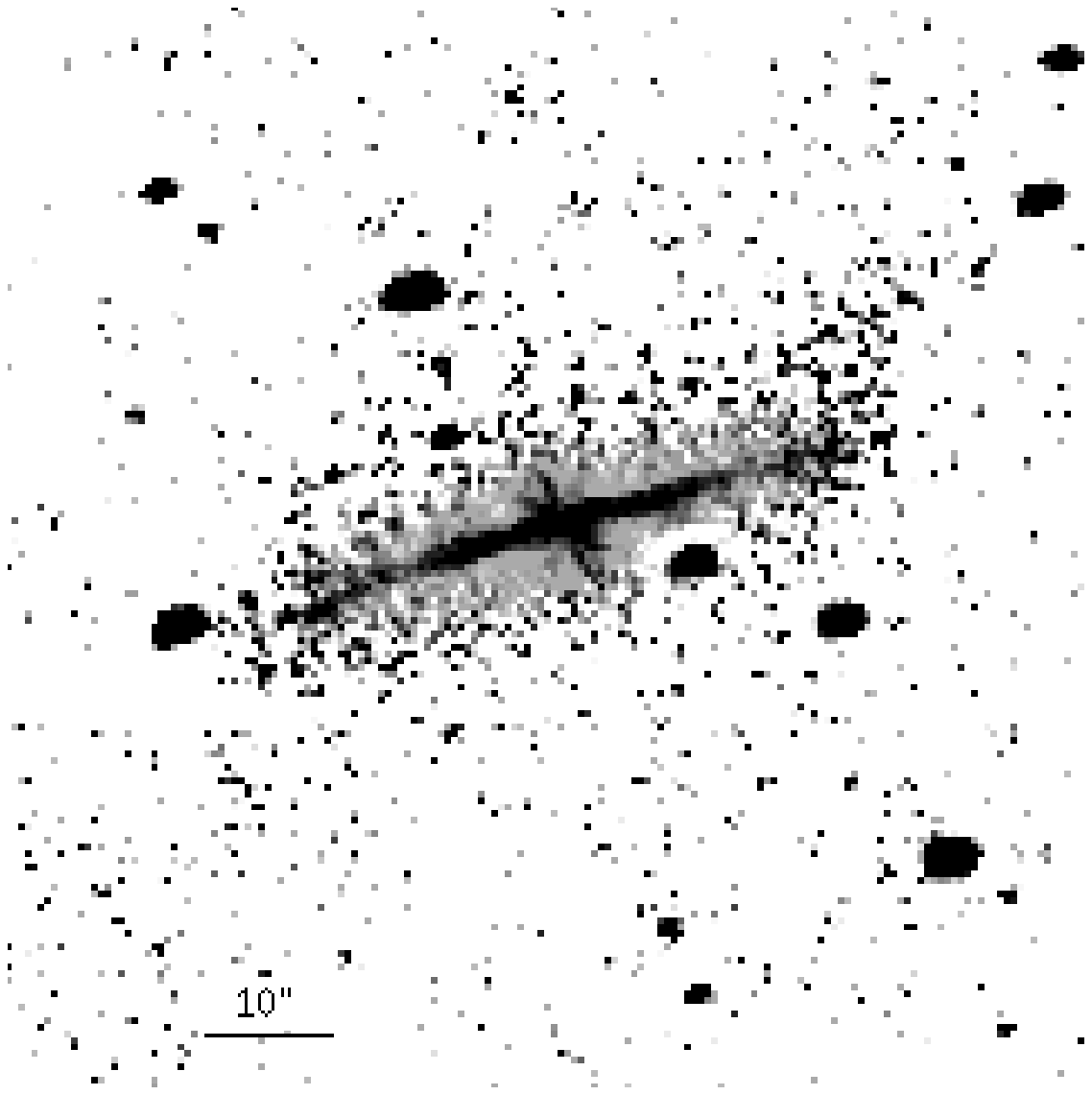} 
\includegraphics[width=6cm]{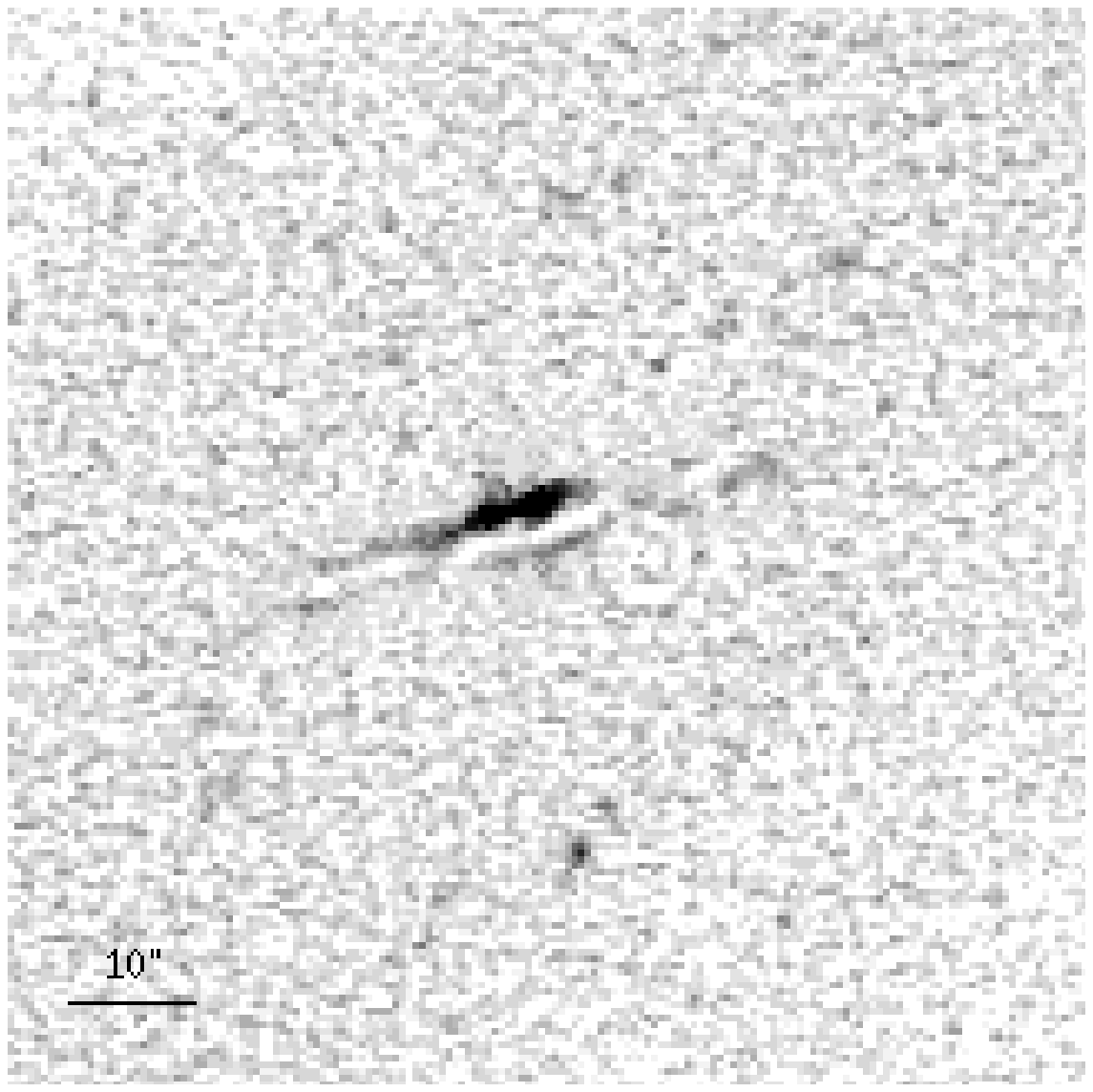} 
\caption{\label{eso235fm} High frequency residual image for 
ESO~235-G58 in the H-band (left panel) and in the B band (right 
panel).  In the B-band image the foreground stars were masked.  The 
darker regions are characterized by a value greater than 1, and 
lighter regions by a value less than 1.  North is at the top and East 
is to the left.} 
\end{figure*}

The NIR unsharp masked images\footnote{The un-sharp masked images are
obtained as ratios between the co-added galaxy frame and its median
filtered image, which is the result of a median filtered image,
computed with the FMEDIAN package in IRAF, where each original pixel
value is replaced with the median value in a sliding rectangular
window. After several tests, a window size of $7 \times 7$ pixels was
chosen because it provides the optimum enhancement of the galaxy inner
structures.}  (see Fig.\ref{eso235fm} for B and H band) confirm that
the central component in ESO~235-G58 is an edge-on disk galaxy: in the
NIR unsharp masked images, a luminous nearly edge-on structure is
clearly identified, along the major axis of the galaxy, (see left
panel of Fig.\ref{eso235fm} for the H band); this feature is then seen
in absorption in the optical images (see right panel of
Fig.\ref{eso235fm} for the B band). It confirms that the disk is not
completely edge-on because the dust lane does not split the bulge into
two equal halves. Furthermore, a ``filamentary'' structure is visible
in the H band, perpendicular to the disk and aligned with the apparent
minor axis of the galaxy, extending approximately $10''$. A
similar feature was also observed in the polar ring galaxy ESO~603-G21
(Iodice et al. 2002b; Reshetnikov et al. 2002; Arnaboldi et al. 1995),
where rotation was detected along a axis coincident with this filament,
suggesting that it is a small edge-on disk. In the case of
ESO~235-G58 we lack any kinematical data, thus no conclusions can
be reached on the nature of this structure perpendicular to the main
disk.
   
\section{Photometry: light and color distribution}\label{eso235_phot}   
   
\begin{figure}   
\centering   
\includegraphics[width=10cm]{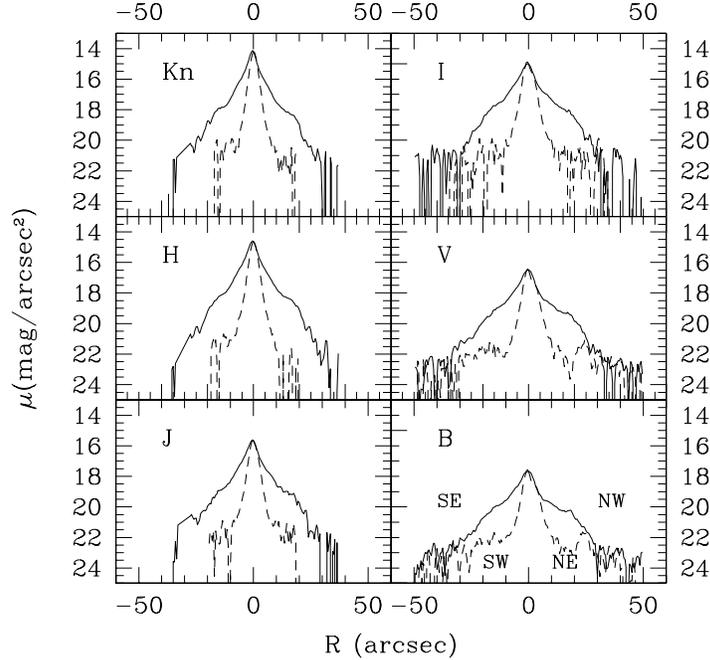} 
\caption{\label{eso235_prof} NIR (left panels) and optical (right profiles)    
surface brightness profiles along the major, $P.A.=106^\circ$, (continuous   
line) and minor axis (dashed line) of the central galaxy.}   
\end{figure}

The surface brightness profiles were derived along the major 
($P.A.=106^\circ$) and minor axis of the central galaxy, in the NIR 
and optical bands, and they are shown in Fig.\ref{eso235_prof}. The 
dust lane perturbs the optical minor axis profiles, making 
them slightly asymmetric.
The H and Kn band light profiles, in 
contrast, are not strongly influenced by the dust absorption. The 
analysis of the NIR light profiles, along the minor axis, shows that 
the bulge component extends out to $10''$, at most; at large distances 
from the center, the additional light along this axis is related to 
the ring, which is brighter in the B and V bands. 
   
From B to J bands, the light distribution along the NW direction, in
the radial interval $5'' < R < 20''$, is more luminous than the
corresponding regions along the SE direction: this excess is more
likely related to the NW regions of the ring, identified also in the
NIR bands as a bright elongation of the central galaxy disk in this
direction (Sec.\ref{eso235_morph} and Fig.\ref{eso235h}).  The change
in slope, observed at $R \sim 20''$ from the center, in both
directions and in all bands, can not be caused by dust absorption
alone, because it is clearly detected also in H and Kn bands, where
this perturbation is reduced to a minimum. This feature in the light
profiles is observed in the same range of radii where the twisting of
the P.A. profile is also detected (Fig.\ref{eso235h}): they are
consistent with the interpretation of ring light superposed to the
galaxy light along the line of sight.
   
\begin{figure}   
\centering   
\includegraphics[width=10cm]{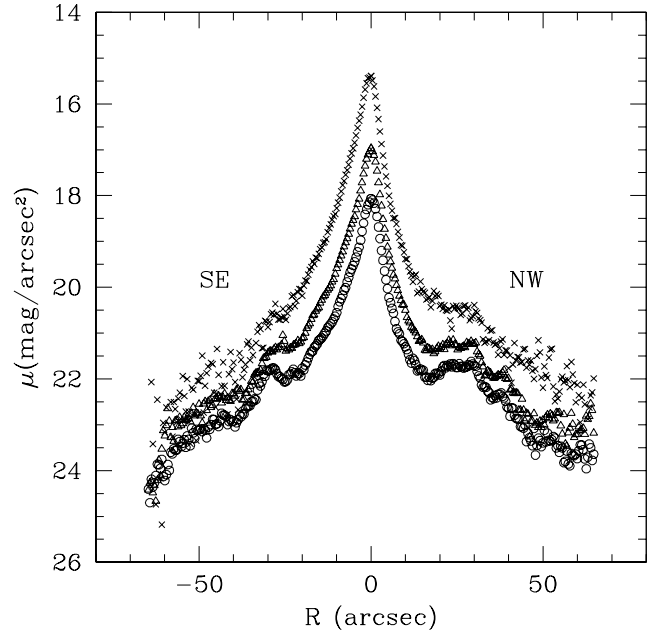} 
\caption{\label{eso235_pr} Surface brightness profiles along the ring major    
axis, $P.A.=146^\circ$, in B (open circles), V (open triangles) and I bands    
(crosses).}   
\end{figure}

The surface brightness profiles for the ring structure, shown in 
Fig.\ref{eso235_pr}, are obtained in the optical bands only, since 
this component is almost undetectable in the NIR bands, as described 
in Sec.\ref{eso235_morph}. In order to cover the whole ring extension, 
including the spiral arms, the surface brightness profiles for this 
component are the results of an average of 40 profiles extracted in a 
cone, centered on the host galaxy and 10 degrees wide from the position 
angle of the ring major axis ($P.A.=146^\circ$, by Buta and Crocker, 
1993). On average, the ring light distribution seems to have an 
exponential profile; the several bumps, which strongly perturb the B 
and V profiles, are caused by star forming regions and dust in the 
ring structure (Fig.\ref{eso235B}). 
   
\begin{figure*}   
\includegraphics[width=6cm]{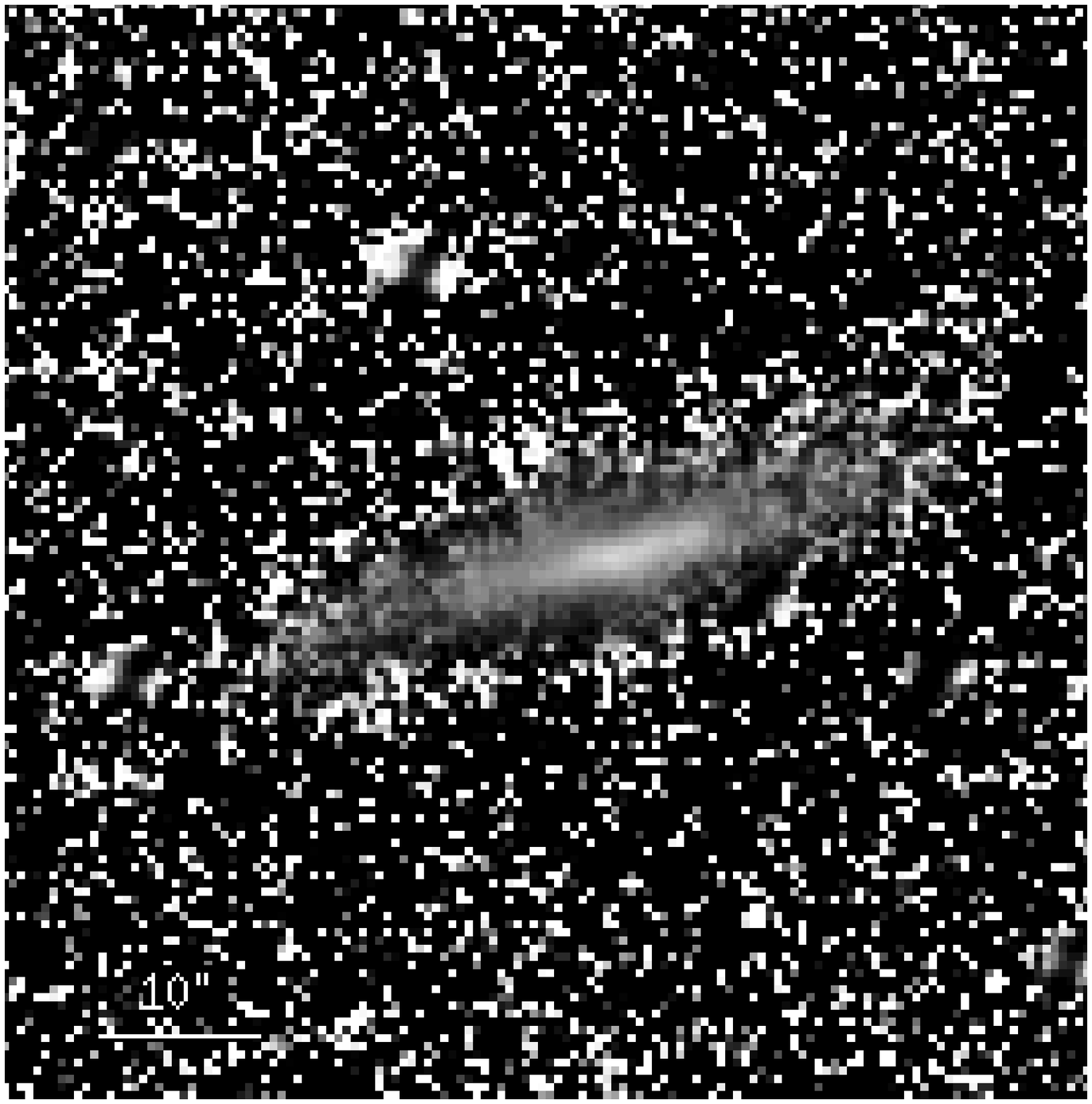} 
\includegraphics[width=6cm]{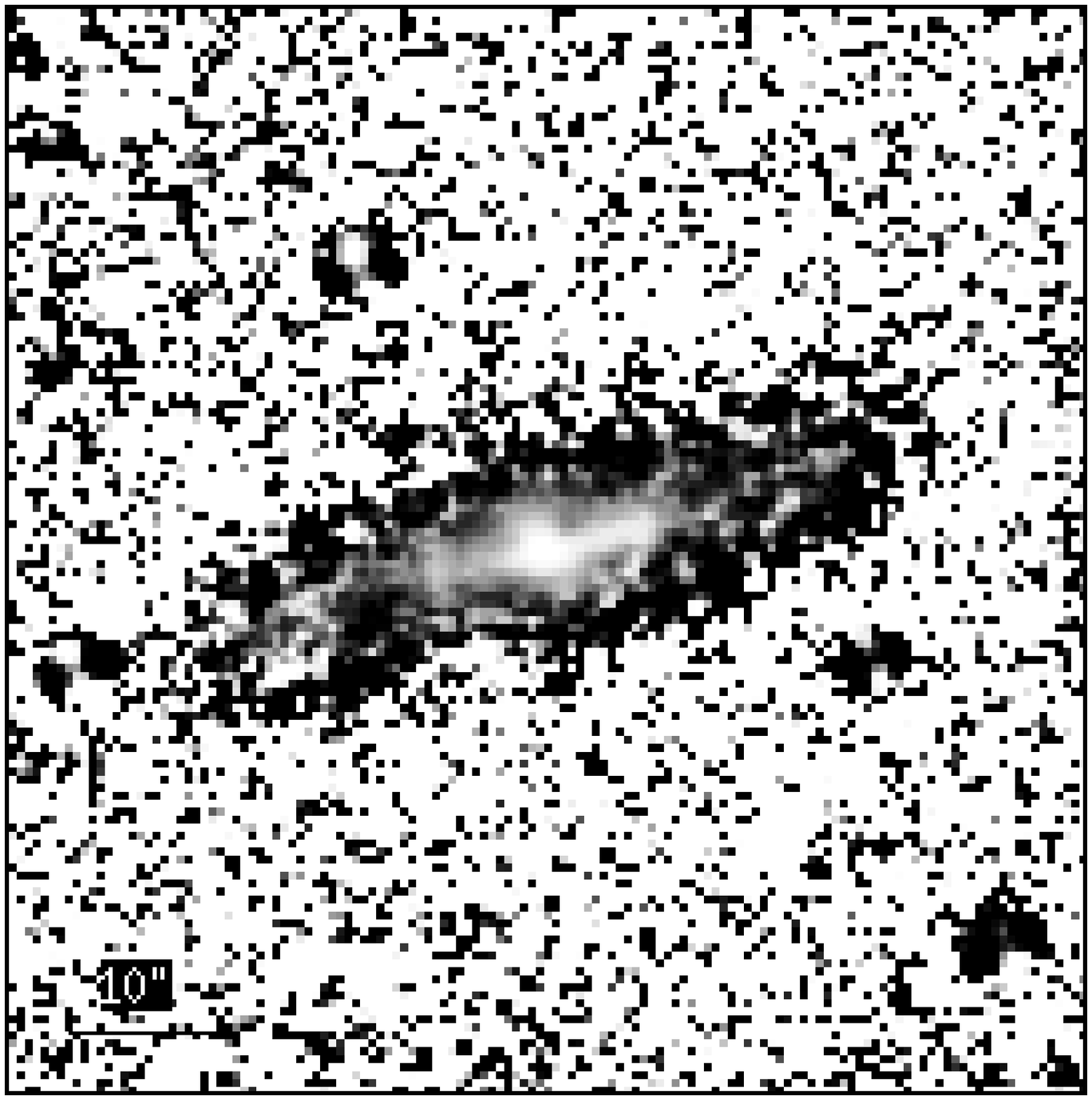} 
\caption{\label{eso235_mappe} J-H (left panel) and H-K (right panel) color    
map for ESO~235-G58. Darker regions correspond to bluer colors and lighter    
regions correspond to redder colors.    
North is at the top and East is to the left.}   
\end{figure*}

\begin{figure}   
\centering   
\includegraphics[width=7cm]{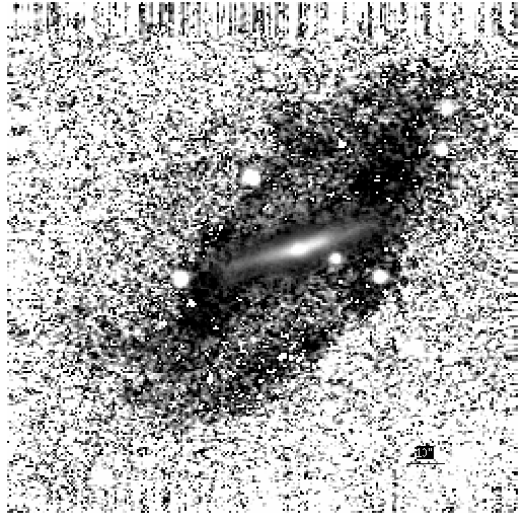} 
\caption{\label{eso235_bk} B-K color map for    
ESO~235-G58. Darker regions correspond to bluer colors and lighter regions    
correspond to redder colors. North is at the top and East is to the left.}   
\end{figure}

The J-H, H-K color maps (Fig.\ref{eso235_mappe}) show that the nuclear 
regions of the galaxy, inside $10''$ from the center, are redder than 
the outer ones: they correspond to the bulge component, whose 
extension has been previously estimated to be about $10''$.  The B-K 
color map (Fig.\ref{eso235_bk}) shows even more details of the 
structure of ESO~235-G58: the very red bulge is again seen, and the 
dust lane along the central object major axis is much redder than its 
surroundings. The very blue regions, all around the central galaxy, 
correspond to the ring structures, which appear more face-on with 
respect to the central object. 
   
\section{Dust properties}\label{eso235_dust}   
   
\begin{figure*}   
\resizebox{\hsize}{!}{   
\includegraphics[width=6cm]{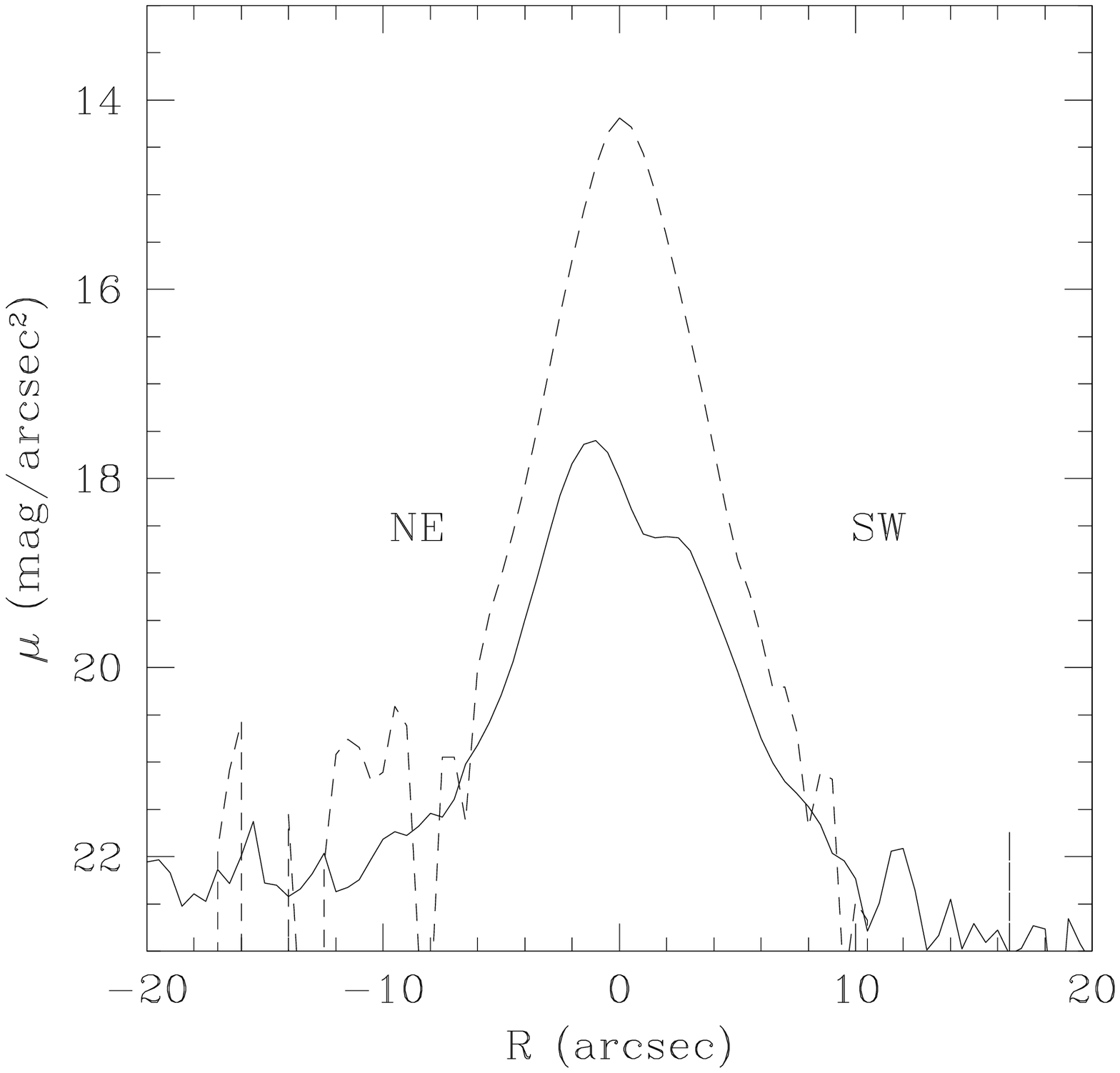} 
\includegraphics[width=6cm]{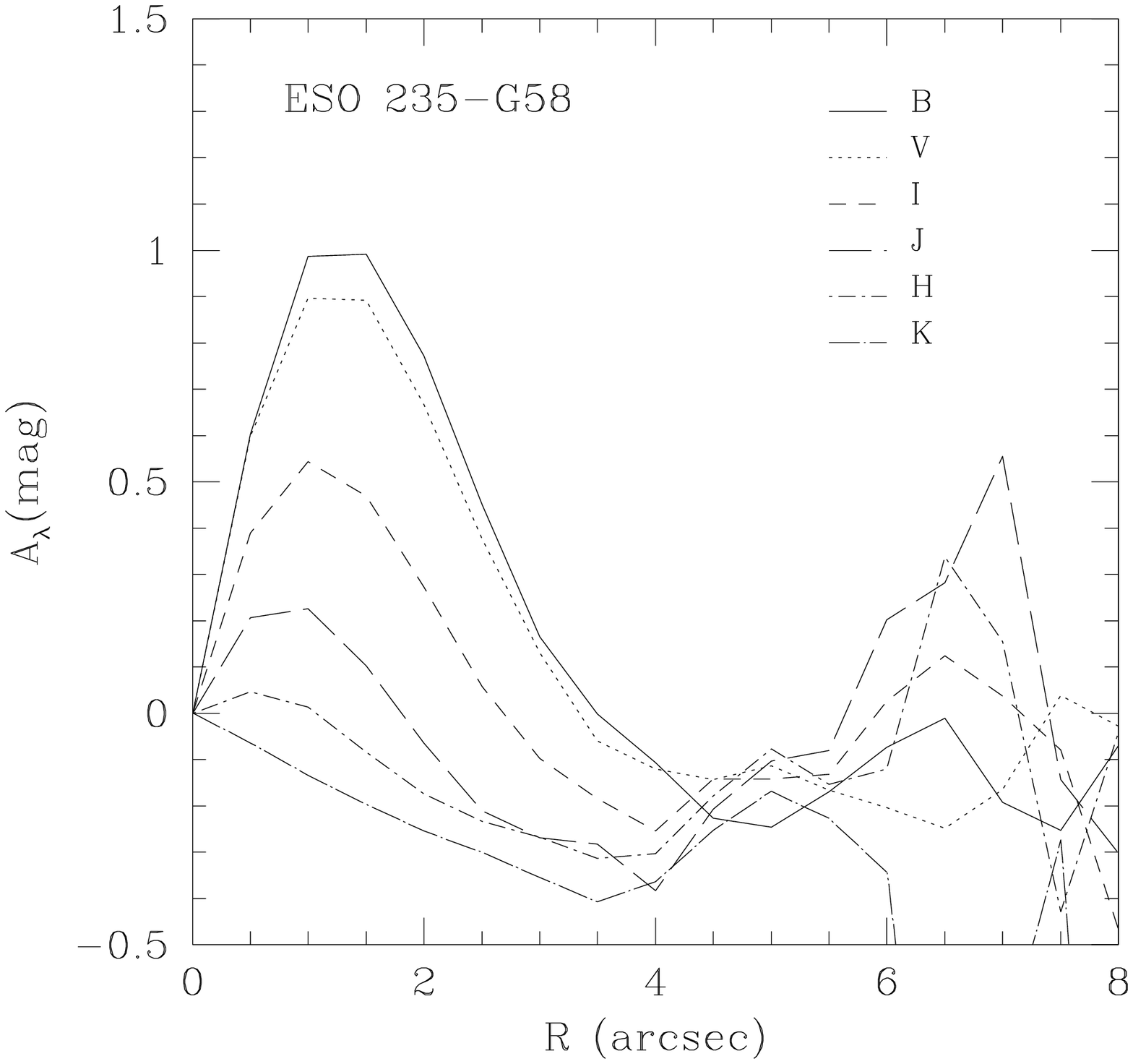} 
\includegraphics[width=6cm]{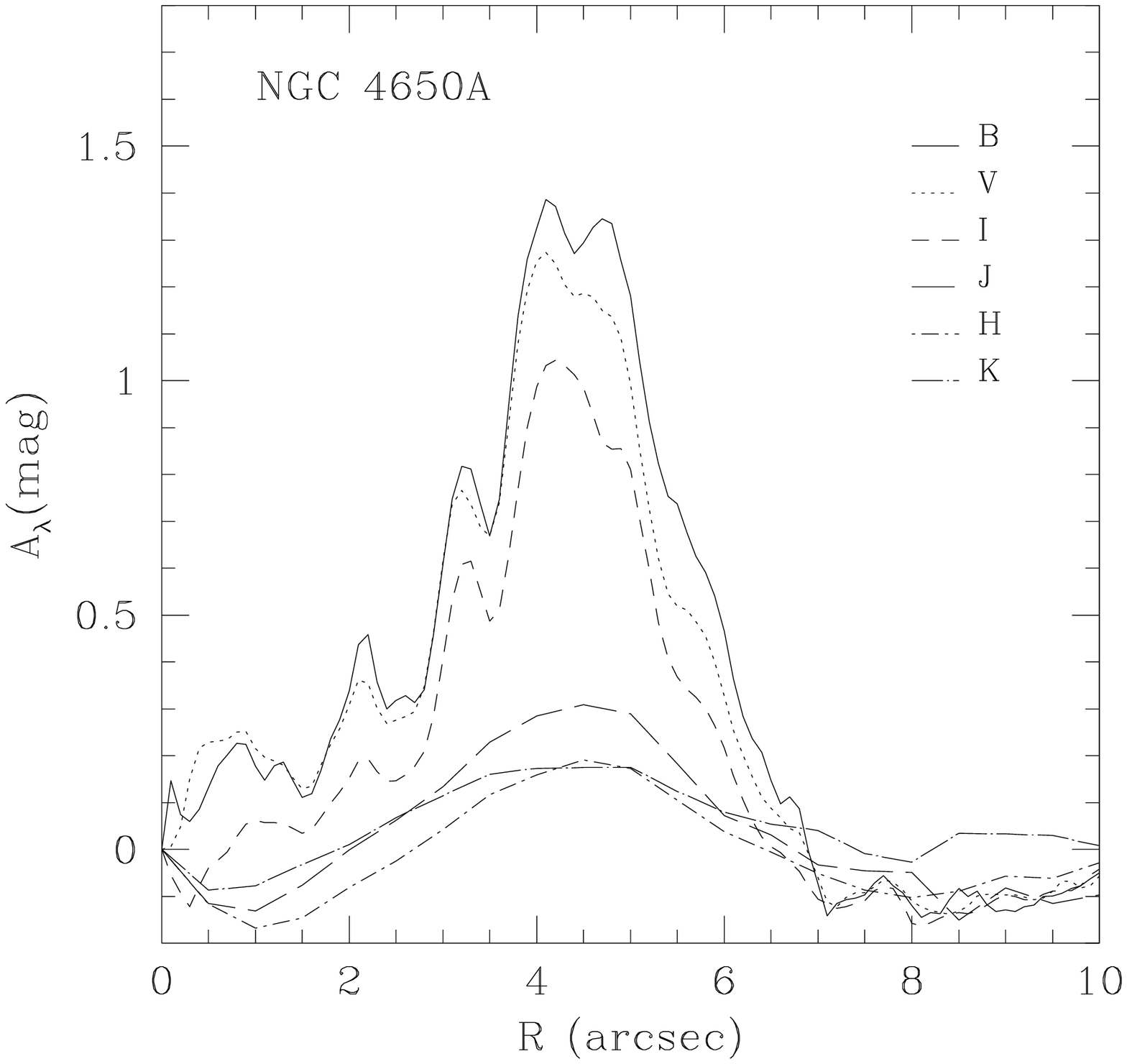} 
}   
\caption{\label{dust_al} Left panel - Surface brightness profiles along the    
minor axis of the inner regions of ESO~235-G58 in B (continuous line)
and in K band (dashed line). Middle panel - The extinction profile
$A_{\lambda}$ observed for the dust lane in ESO~235-G58, in the B, V,
I, J, H and K bands.  Left panel - The extinction profile
$A_{\lambda}$ observed for the dust lane in the polar ring galaxy NGC
4650A.}
\end{figure*}

The central galaxy of ESO~235-G58 is characterized by a relatively 
high degree of symmetry and a high inclination: these properties make 
this object suitable for the study of the dust lane, which runs 
parallel to the major axis out to $17''$ from the center of the galaxy 
(see Fig.\ref{eso235B}). In order to estimate the extinction law in 
the dust lane, we adopted the same analysis used by Knapen and 
collaborators (Knapen et al.\ 1991) for the Sombrero galaxy (NGC 
4594): in each band, we compare the surface brightness distribution of 
regions perpendicular to the dust lane, i.e. along the minor axis, 
which are at the same distance from the galaxy center. We then 
subtracted the unobscured part of the profile, i.e. the northern side, 
from its obscured counterpart, the southern side, in order to obtain 
the {\it absorption profile} defined by $$A_{\lambda} (R)=-2.5log 
\frac{I_{obs}(\lambda)}{I_{true}(\lambda)}$$ 
   
\noindent   
where $I_{obs}$ is the observed intensity at a given point in the dust 
lane and $I_{true}$ is the intensity relative to the starlight with no 
dust obscuration. For this analysis, the exact center of the galaxy is 
determined from the K band image. The minor axis profile relative to 
the inner regions of the galaxy are shown in the left panel of 
Fig.\ref{dust_al}: the dust absorption strongly perturbs the B band 
profile from the center to about $4''$ radius on the SW side, while 
this effect is almost absent in the K band profile. The absorption 
profiles relative to each band are plotted in the middle panel of 
Fig.\ref{dust_al}: the dust absorption is significant in the J band 
also, and becomes weaker in the H band. The absorption profile in the 
K band shows that there is no absorption caused by the dust lane in this 
region, but on the contrary it suggests that the SW side of the 
profile is brighter (about 0.4 mag) than the correspondent NE side. 
This is probably due to the twisting of the isophotes observed from 
the center to about $4''$ (see Fig.\ref{eso235h}) and ``hidden'' by 
the dust absorption in the other bands.\\ By using standard linear 
regression, we have obtained the values of the ratio $A_{\lambda}/A_V$ 
presented in Tab.\ref{Alambda}. These values are compared with those 
typical for our Galaxy (Rieke \& Lebofsky 1985), for the Sombrero 
galaxy (Knapen et al. 1991) and for the polar ring galaxy NGC 4650A 
(Iodice et al. 2002a), for which we have derived the absorption 
coefficient profiles\footnote{We have compared the surface brightness 
profiles along the major axis of the host galaxy, where the dust in 
the polar ring obscures the SW regions (see Iodice et al. 2002a for 
details).}  shown in Fig.\ref{dust_al} (right panel).  Taking into 
account the errors, the values of $A_{\lambda}/A_V$ for ESO~235-G58 
are consistent with those derived for the Sombrero galaxy and for NGC 
4650A and for those relative to our Galaxy only in the near-IR, JHK 
bands. Major differences are observed between the $A_{\lambda}/A_V$ 
for our Galaxy and those for the other three galaxies in the optical 
bands. As suggested by Knapen et al. (1991), such differences do not 
necessarily imply that the properties of the dust in our Galaxy are 
different from the other galaxies. Some fraction of the light in the 
dust lane does not originate behind the dust. In the simple models 
developed to describe this phenomenon (Walterbos \& Kennicutt 1988) 
the absorption profile $A_{\lambda}$ is also a function of the optical 
depth of the dust, which becomes progressively smaller toward longer 
wavelengths, implying an even smaller difference between the 
$A_{\lambda}/A_V$ ratio for our Galaxy and for other galaxies. 
   
\begin{table}   
\centering   
\caption[]{Extinction ratios $A_{\lambda}/A_V$ for the dust lane in ESO~235-G58, for the Sombrero galaxy (NGC 4594), for the polar ring galaxy NGC 4650A and for our Galaxy, in several bands.}   
\label{Alambda}   
\begin{tabular}{ccccc}   
\hline\hline   
Band & ESO~235-G58& NGC 4594 & NGC 4650A & our Galaxy \\    
\hline   
$A_B/A_V$ & 1.13 $\pm$0.04& 1.20 & 1.11 $\pm$0.01& 1.324\\   
$A_V/A_V$ & 1.00& 1.00 & 1.00 & 1.00\\   
$A_I/A_V$ & 0.67 $\pm$0.06& 0.52 & 0.83 $\pm$0.01& 0.482\\   
$A_J/A_V$ & 0.37 $\pm$0.11& 0.31 & 0.31 $\pm$0.03& 0.282\\   
$A_H/A_V$ & 0.15 $\pm$0.09& 0.19 & 0.20 $\pm$0.03& 0.175\\   
$A_{K_n}/A_V$& 0.10 $\pm$0.06& 0.09 & 0.14 $\pm$0.03& 0.112\\   
\hline   
\end{tabular}   
\end{table}   
   
\section{Integrated magnitudes and colors}\label{eso235_col}   
   
We wish to analyze the integrated colors derived for the main 
components (the central galaxy and ring) in ESO~235-G58, in order to 
compare them with those of standard morphological galaxy types. To 
this aim, the integrated magnitudes are computed in several areas 
which cover the two components and are selected differently in the 
optical compared to the NIR bands, because of the different 
morphologies in these bands (Sec.\ref{eso235_morph}).   
A detailed discussion about the photometric 
error estimate, which takes into account both photon statistics and 
background fluctuations, was presented by Iodice et al. (2002b). In 
the optical bands they are about $3\%$, whereas in the NIR they are 
about $10\%$.  The integrated magnitudes and colors derived in each 
area are then corrected for the galactic extinction within the Milky 
Way, by using the values for the absorption coefficient in the B band 
($A_B$) and the color excess $E(B-V)$ derived from Schlegel et 
al. (1998). The absorption coefficients $A_\lambda$ are derived for 
each band, by adopting $R_V=A_V/E(B-V)=3.1$ from Cardelli et 
al. (1989).\\ 
   
\subsection{Optical colors}\label{eso235_col_opt}   
\begin{figure}   
\centering   
\includegraphics[width=7cm]{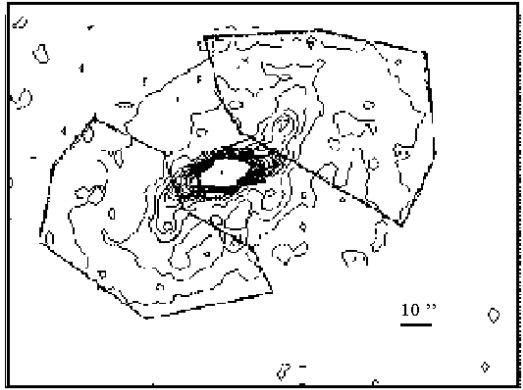} 
\caption{\label{eso235_polyb} ESO~235-G58 contour plot in B band    
(continuous line) plus the 3 polygons limiting the different areas  
where the integrated magnitudes are computed. The polygons enclosing  
the central galaxy (not clearly visible on the figure) is almost  
coincident with an ellipse whose major axis is about $20''$. North is  
at the top and East is to the left.}  
\end{figure}

\begin{figure}   
\centering   
\includegraphics[width=7cm] {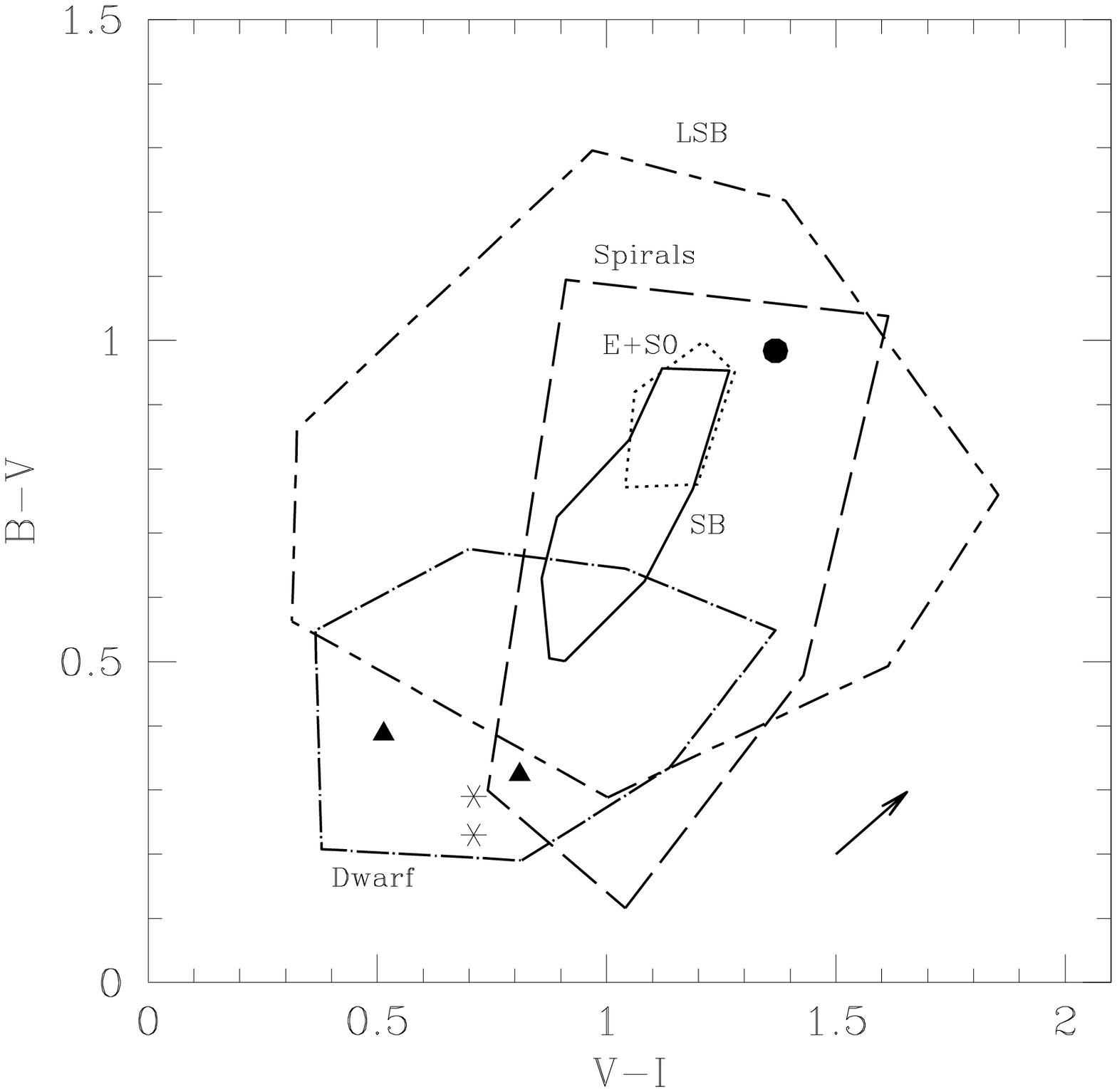} 
\caption{\label{eso235_bvi} B-V vs. V-I color diagram for the central galaxy   
(filled circle) and ring (filled triangles) in ESO~235-G58. Asterisks 
are the optical colors for the polar structure in NGC4650A.  The 
dotted contour limits the region where the integrated colors of Es and 
S0s are found; the long-dashed contour limits the integrated colors of 
spirals; the dashed-dotted contour identifies the integrated colors of 
the dwarf galaxies, the long dashed - short dashed contour identifies 
the integrated colors of LSB galaxies and the continuous contour 
limits the integrated colors of barred galaxies with outer rings.  The 
arrow, in the lower right corner, indicates the reddening vector for 
galactic dust and the screen model approximation.} 
\end{figure}

In the B,V, and I bands, the areas where the integrated magnitudes are 
computed are selected as follows: one is coincident with the central 
galaxy, and two areas are for the ring structure, along the SE and NW 
directions (see Fig.\ref{eso235_polyb}). The polygons are determined 
in the B band, where the ring component is more luminous and extended, 
and they are used for the V and I images, previously registered to the B 
image. The integrated magnitudes and colors, derived for each area, 
are listed in Tab.\ref{eso235_opt}. The B-V vs. V-I colors for the 
central galaxy and ring (Fig.\ref{eso235_bvi}) are compared with those 
of (1) standard early-type galaxies (Michard \& Poulain 2000), (2) 
spiral galaxies (de Jong \& van der Kruit, 1994), (3) dwarf galaxies 
(Makarova 1999), (4) LSB galaxies (O' Neil et al. 1997; Bell et 
al. 2000), and (5) barred galaxies with outer rings and pseudo-rings 
(Buta \& Crocker, 1992). The colors of the central galaxy (which are 
redder\footnote{This is probably due to a different area chosen in the 
present work to estimate colors.} than those derived by Buta \& 
Crocker 1993), taking into account the dust reddening, are within the range of colors for spiral and LSB galaxies. 

The outer ring is much bluer than the central galaxy and the colors we 
find are consistent with the values derived by Buta \& Crocker 
(1993). The optical colors of this component are similar to those of 
dwarf galaxies and the polar structure in the polar ring galaxy 
NGC4650A (Iodice et al. 2002a), as shown in Fig.~\ref{eso235_bvi}. The 
strong color gradient between the central galaxy and ring cannot be 
the effect of dust reddening only, since the reddening vector cannot 
account for the whole color difference. Therefore it implies a 
difference in stellar populations.\\ 
   
\subsection{NIR colors}\label{eso235_col_nir}   
\begin{figure}   
\centering   
\includegraphics[width=7cm]{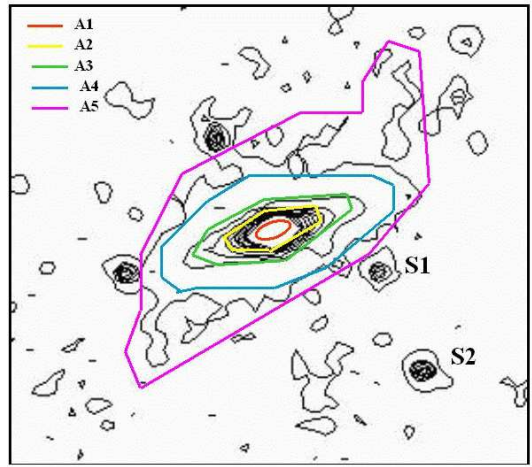} 
\caption{\label{eso235_polyk} ESO~235-G58 contour plot in Kn band (continuous 
line) plus the concentric five polygons limiting the 
areas where the integrated magnitudes are computed. The distance 
between the two stars (S1 and S2) is about $20''$. North is at the top 
and East is to the left.} 
\end{figure}

\begin{figure*}   
\includegraphics[width=7cm]{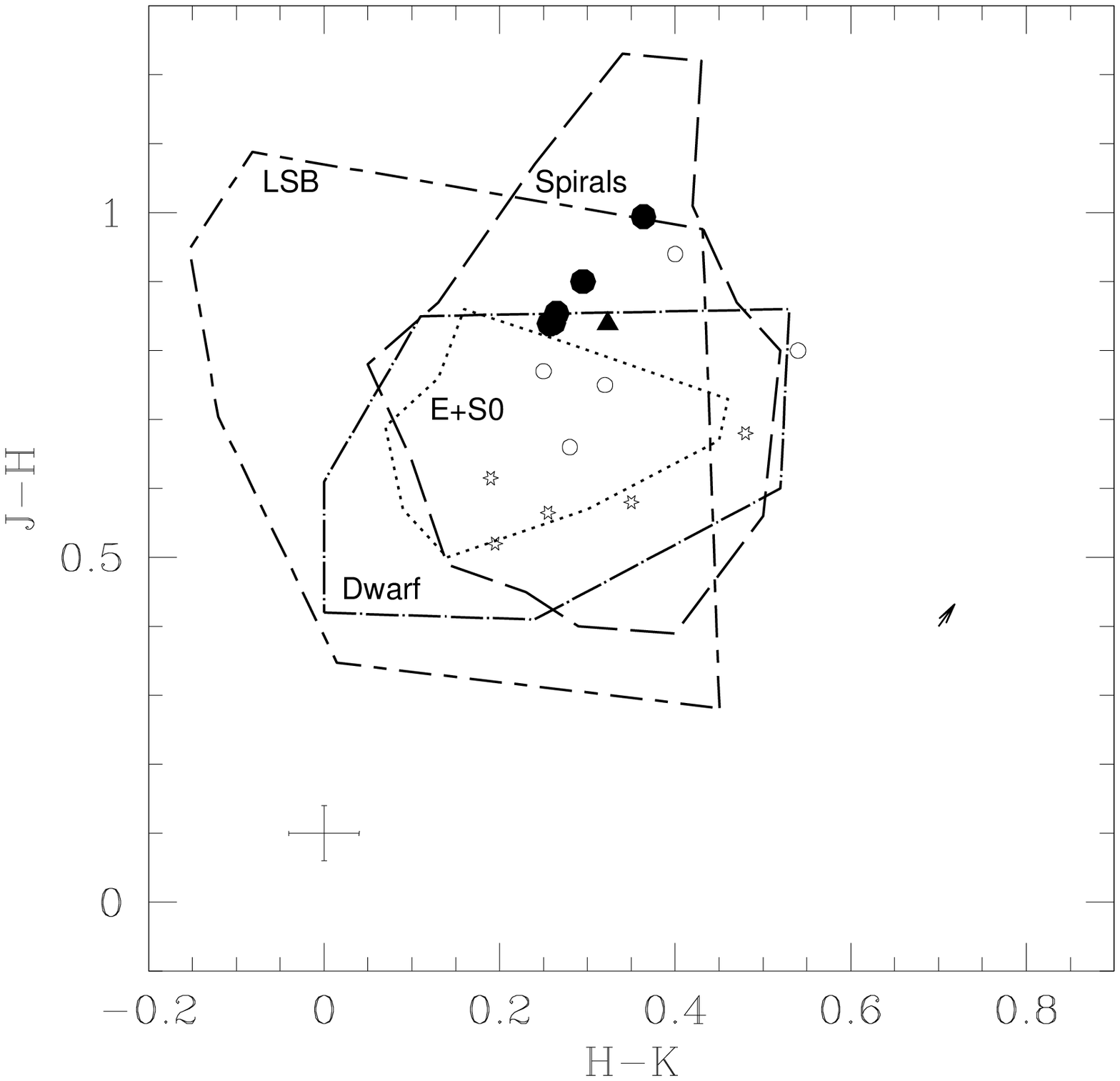} 
\includegraphics[width=7cm]{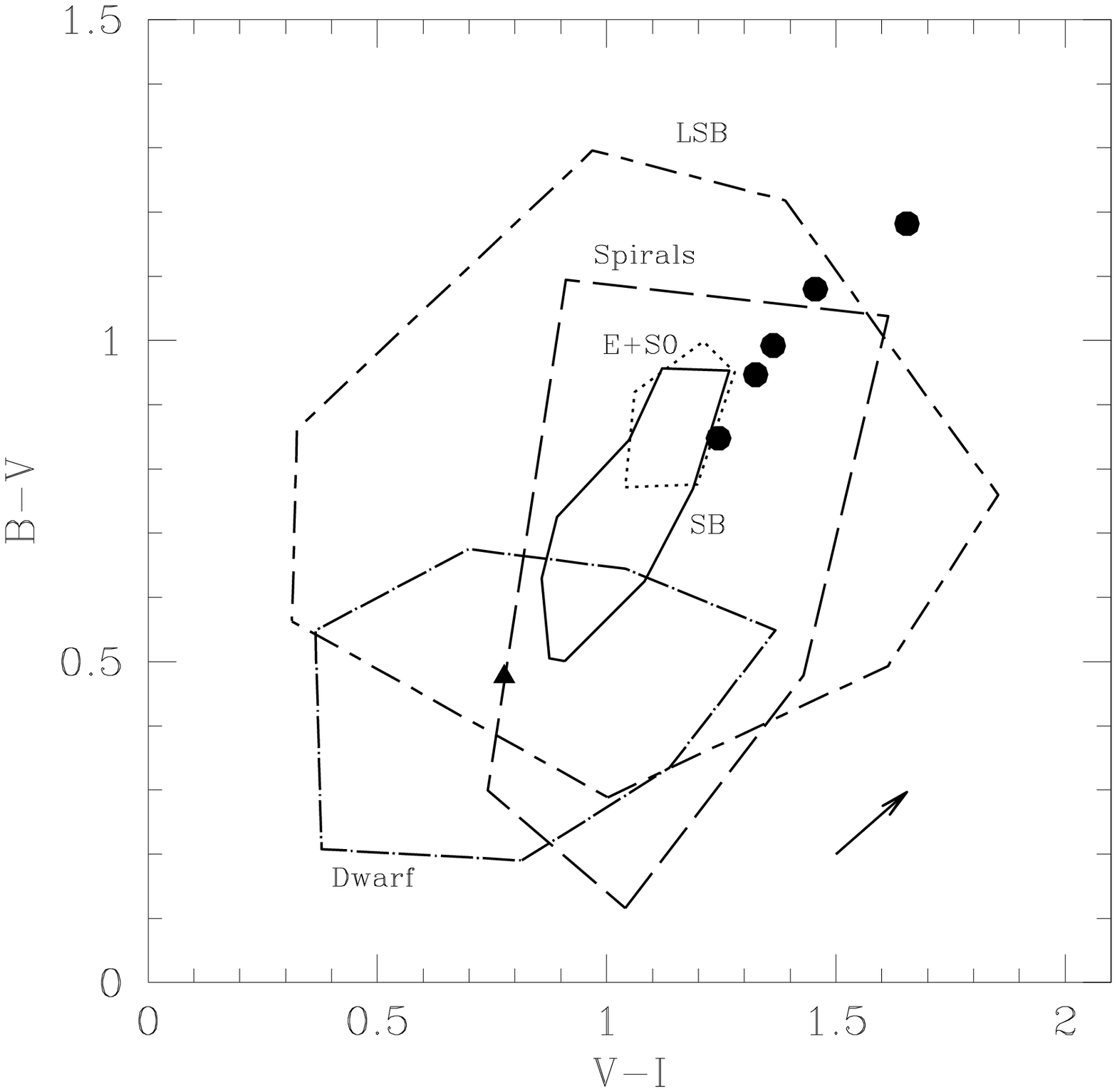} 
\caption{\label{eso235_jhk}    
J-H vs. H-K (left panel) and B-V vs. V-I (right panel) color diagram 
for the central galaxy (filled circles) and ring (filled triangles) in 
ESO~235-G58.  Open circle indicates the colors of the nuclear regions 
in PRGs, and open stars indicate the average colors of the host galaxy 
in PRGs (data are from Iodice et al. 2002b,c). The dotted contour 
limits the region where the integrated colors of Es and S0s are found; 
the long-dashed contour limits the integrated colors of spirals; the 
dashed-dotted contour identifies the integrated colors of the dwarf 
galaxies and the long dashed - short dashed contour identifies the 
integrated colors of LSB galaxies and the continuous contour limits 
the integrated colors of barred galaxies with outer rings. The arrow, 
in the lower right corner, indicates the reddening vector for galactic 
dust and the screen model approximation.} 
\end{figure*}

The study of the colors in the NIR concerns mainly the central galaxy, 
since the ring is hardly detected in these bands (see 
Sec.\ref{eso235_morph}) and it is quite difficult to define any areas 
where magnitudes and colors can be reliably estimated. Five 
polygons\footnote{These polygons are different from those defined for 
the optical images, described in Sec.\ref{eso235_col_opt}.}, with the 
same center and larger areas, were defined starting from the galaxy 
contour in the Kn band. The largest polygon is determined from the 
outer isophote in the Kn band associated with the galaxy 
(Fig.\ref{eso235_polyk}). The integrated magnitudes and colors for the 
ring component were estimated from the difference between fluxes in 
the last two polygons. This choice is suggested by the excess blue 
light around the central galaxy in the NIR color maps 
(Fig.\ref{eso235_mappe}). The five polygons are used both for J and H 
bands and for the optical bands, after the images were registered and 
scaled to the Kn image. The integrated magnitudes and colors, 
corresponding to each area, are listed in Tab.\ref{eso235_nir}. 
Fig.\ref{eso235_jhk} plots the J-H vs. H-K (left panel) and B-V 
vs. V-I (right panel) color diagrams. In both the NIR and in the 
optical, colors become bluer as area increases. In both color 
diagrams, on average, the central galaxy has colors very similar to 
those of spiral galaxies; the nuclear region of this component is 
characterized by the reddest colors. Furthermore, in the optical 
bands, the central galaxy is on average redder than the ringed barred 
galaxies. The inner regions of the ring are much bluer than the 
central galaxy, as already found for the colors of the non-scaled 
optical images (see previous section). In the NIR, on the contrary, 
the central galaxy and ring have similar colors. Since the ring 
component is very faint in these bands (as also suggested by the 
constant values of the colors in the last three areas, see 
Tab.\ref{eso235_nir}) the corresponding colors may be influenced by 
background noise. 
    
How do the NIR colors of ESO~235-G58 compare with typical colors of 
Polar Ring Galaxies?  A common characteristic of ESO~235-G58 and PRGs 
is the very red nucleus with respect to the outer regions (see 
Fig.\ref{eso235_jhk}). The central galaxy of ESO~235-G58 has redder 
colors than the average PRG host galaxy (Iodice et al.\ 2002b,c).

\begin{table}   
\centering   
\caption[]{Integrated magnitudes and colors, in the optical bands, for the    
central galaxy and ring in ESO~235-G58.}   
\label{eso235_opt}   
\begin{tabular}{ccccc}   
\hline\hline   
Component & $m_B$ & $M_B$ & B-V & V-I\\   
\hline   
GALAXY & 16.27 & -17.6 & 0.98 & 1.37\\   
RING (NW)& 16.73 & -17.2 & 0.32 & 0.81\\   
RING (SE)& 16.93 & -17.0 & 0.39 & 0.51\\   
\hline   
\end{tabular}   
\end{table}   
   
\begin{table}   
\centering   
\caption[]{Integrated magnitudes and colors of different regions in    
ESO~235-G58. In the second column are listed the largest distances from the    
center (in arcsec) reached by each polygon.}   
\label{eso235_nir}    
\begin{tabular}{cccccccccc}   
\hline\hline   
Region & a (arcsec) & $m_B$ (mag) & $m_J$ (mag)& B-V& V-I& B-H& J-K& J-H& H-K\\   
\hline   
A1 & 2.7 & 18.51 & 14.72 & 1.18& 1.65& 4.78& 1.36& 0.99& 0.36\\                   
A2 & 10  & 16.85 & 13.59 & 1.08& 1.45& 4.16& 1.19& 0.90& 0.29\\                   
A3 & 18.2& 16.22 & 13.23 & 0.99& 1.36& 3.84& 1.12& 0.85& 0.26\\                   
A4 & 22.7& 16.01 & 13.13 & 0.95& 1.32& 3.72& 1.10& 0.84& 0.26\\                   
A5 & 40  & 15.7  & 13.05 & 0.85& 1.24& 3.49& 1.10& 0.84& 0.26\\                   
RING &   & 17.19 & 15.97 & 0.48& 0.78& 2.06& 1.16& 0.84& 0.32\\                   
\hline   
\end{tabular}   
\end{table}   
   
\section{Age estimate in ESO~235-G58}\label{age_eso235}    
   
We wish to derive an estimate of the stellar population ages in the 
central galaxy and in the apparent inner pseudo-ring ring for this 
peculiar object and to compare them with the typical ages for other 
morphological galaxy types. To this aim, the stellar population 
synthesis model developed by Bruzual \& Charlot (1993), GISSEL ({\it 
Galaxies Isochrone Synthesis Spectral Evolution Library}), was used to 
reproduce the B-H and J-K integrated colors of different regions (see 
Sec.\ref{eso235_col}) in ESO~235-G58. This wavelength range was 
already adopted by Iodice et al. (2002c) to estimate the age of the 
stellar populations in PRGs, since it helped in breaking the 
age-metallicity degeneracy (see also Bothun et al. \cite{Bothun84}). 
   
 
The GISSEL key input parameters are the Initial Mass Function (IMF), 
the Star Formation Rate (SFR), and the metallicity.  In what follows, 
we have assumed that stars form according to the Salpeter 
(\cite{Salp55}) IMF, in the range from $0.1$ to $125 M_\odot$. 
   
\begin{figure}   
\centering   
\includegraphics[width=7cm]{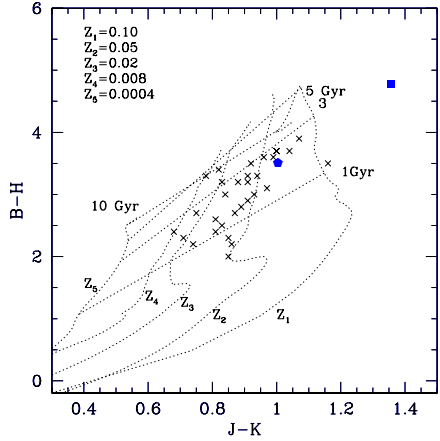} 
\caption{B-H vs. J-K diagram of the evolutionary tracks   
for the stellar synthesis models optimized for the central galaxy in 
ESO~235-G58.  The heavier dotted lines correspond to models with a 
characteristic timescale $\tau=1 Gyr$.  Models are computed for 
different metallicities as shown on this figure.  Light dotted lines 
indicate loci of constant age for the different models; different ages 
are reported on the plot. The filled square and pentagon correspond 
respectively to the nucleus and the outer regions of the central 
galaxy in ESO~235-G58; crosses are for a sample of spiral galaxies 
(Bothun et al., 1984).} 
\label{eso235_age}   
\end{figure}

In the NIR, where the perturbations due to dust are less significant,
the morphology and colors of the central galaxy in ESO~235-G58, are
very similar to a spiral galaxy (like an Sa galaxy), as emphasized in
Sec.\ref{eso235_morph} and in Sec.\ref{eso235_col_nir}. Thus, for this
component, we adopt a star formation history with an exponentially
decreasing rate, $SFR(t)=
\frac{1}{\tau} \exp{(- t/ \tau)}$, with a very small time-scale 
$\tau$, in order to approximate a single burst of star formation. The 
adopted values for the time scale parameter is $\tau=1$ Gyr.  The 
evolutionary tracks corresponding to this model were derived for 
different metallicities,  
which were assumed constant with age, and they are plotted in 
Fig.\ref{eso235_age}, for the B-H and J-K colors. The lines of 
constant age were computed from the evolutionary tracks: for the 
central galaxy in ESO~235-G58, they suggest an age between $1$ to $3$ 
Gyr. At this epoch, we derive a mass-to-light ratio for this component 
of $M/L =1.3$.\\ 
   
As stressed in Sec.\ref{eso235_morph}, the ring structure is too faint
in the NIR bands to derive a realistic age estimate for this
component, by using the B-H and J-K colors. Thus we used only optical
colors (B-V vs. V-I) to derive a characteristic age for the brightest
stellar population.  For this component, a constant star formation
rate was adopted: this choice was mainly suggested by the active star
formation regions observed in the B and V bands (see Fig.\ref{eso235B}
and Buta \& Crocker 1993).  The ring structure seems to be much
younger than the central galaxy: the last episode of star formation is
recent and it may be not older than $10^8$ yr. The mass-to-light ratio
we estimate for this component at this age is $M/L= 0.04$ .
   
The age for the central galaxy of ESO~235-G58 is very similar both to 
the typical age derived for the host galaxy in PRGs (see Iodice et 
al. 2002c) and for spiral galaxies; the ring component could be as 
young as the ring in NGC~4650A (Iodice et al. 2002a) and in 
ESO~603-G21 (Iodice et al. 2002c). 
                          
\section{Study of the light distribution in the central galaxy of ESO~235-G58}   
\label{eso235_g}   
   
In order to derive a quantitative morphology of ESO~235-G58 we 
performed a two-dimensional (2D) model of the central galaxy light 
distribution in the Kn band, where the perturbations by dust 
absorption are weaker.  The 2D model we used to fit the surface 
brightness distribution in ESO~235-G58 is the same adopted to study 
the host galaxy light distribution in PRGs, which is described in 
Iodice et al. 2002b (see also Iodice et al. 2001). The Kn band light 
distribution in ESO~235-G58 was modeled by the superposition of a 
spheroidal central component, whose projected light follows the 
generalized de Vaucouleurs law,  
and an exponential disk.    
The structural parameters which characterized the light distribution of the    
spheroidal component are the {\it effective surface brightness $\mu_{e}$},   
the {\it effective radius $r_{e}$}, the {\it shape parameter n} and    
{\it apparent axial ratio $q_{b}$}; those relative to the disk are    
{\it the central surface brightness $\mu_{0}$}, the {\it scalelength $r_{h}$}   
and  {\it apparent axial ratio  $q_{d}$}.   
   
\begin{figure*}   
\includegraphics[width=7cm]{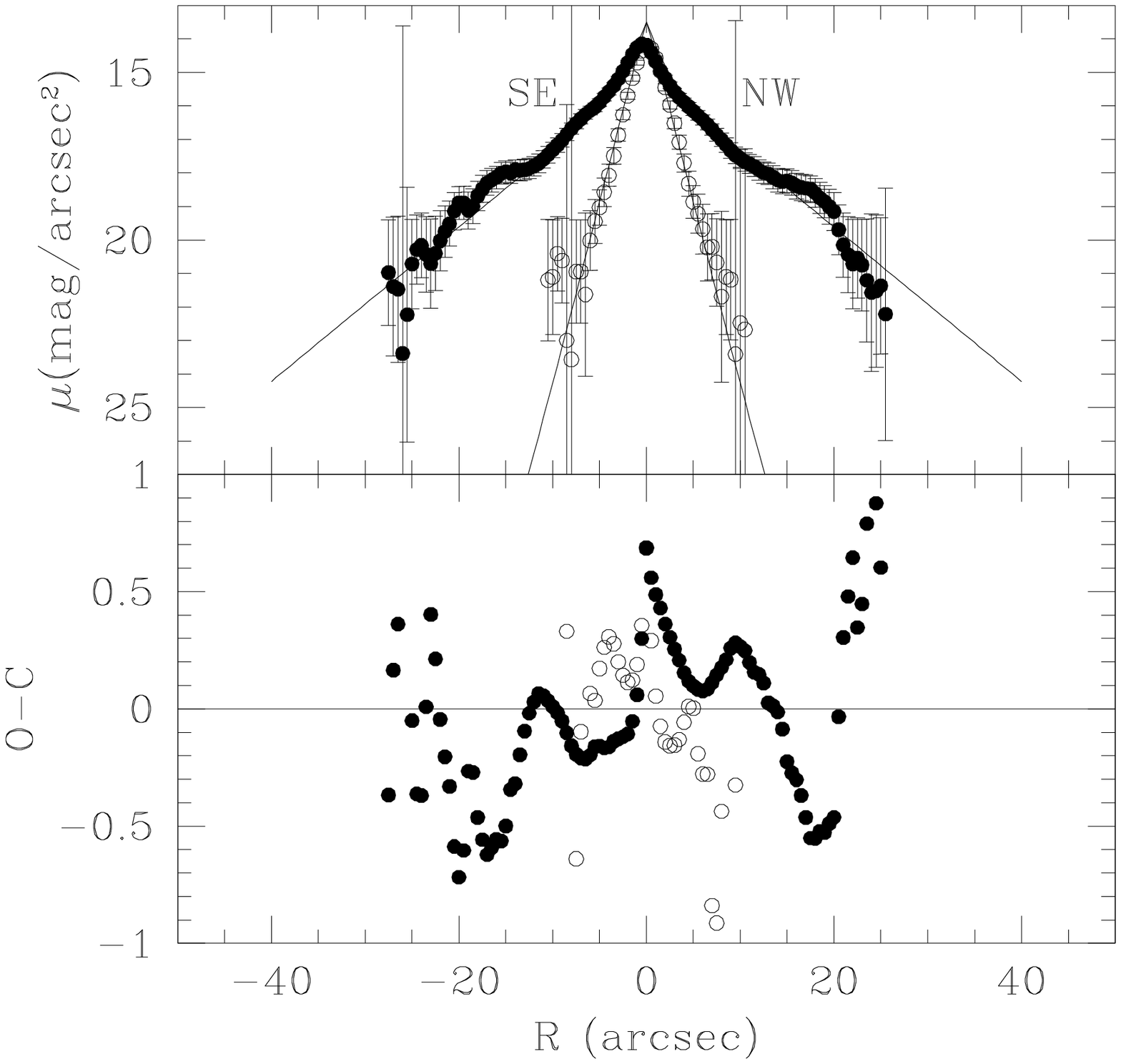} 
\includegraphics[width=6cm]{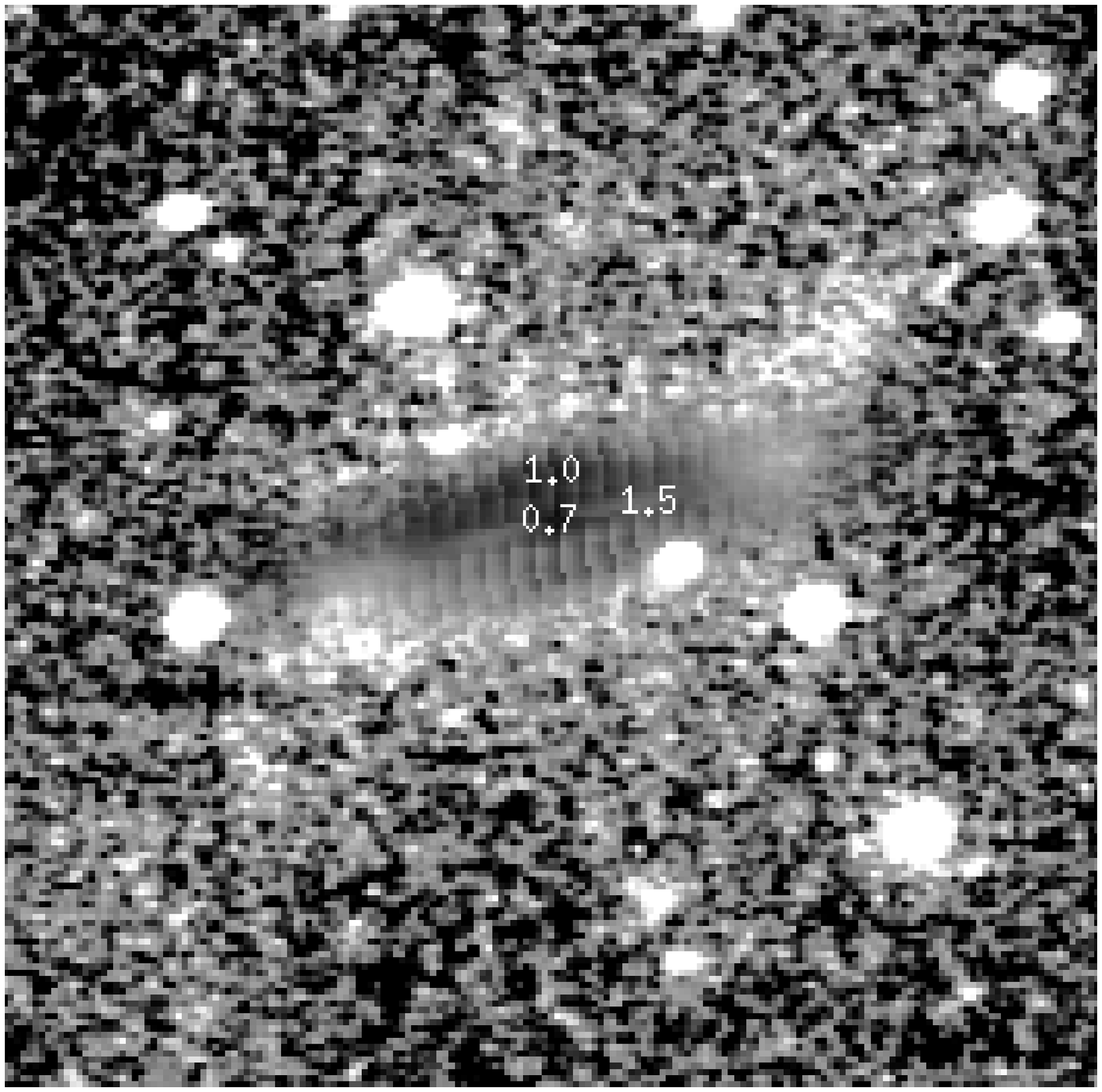} 
\caption{Left panel - 2D fit of the K-band central galaxy light distribution in   
ESO~235-G58.  The observed light profiles along the major (filled
dots), $P.A.=106^\circ$, and minor axis (open dots), $P.A.=196^\circ$,
are compared with those derived by the fit (continuous line), which
was performed in the Kn band. The orientation, reported on the plot,
refers to the major axis. Right panel - Residual image obtained as the
ratio between the whole image of ESO~235-G58 and the 2D model for the
central galaxy in the H band. Units are intensity; whiter colors
correspond to those regions where the galaxy is brighter than the
model, darker colors correspond to those regions where the galaxy is
fainter than the model.  Numbers indicate the value of the ratio at
that particular isophote.  North is up and East is to the left.}
\label{eso235_fit}   
\end{figure*}

The regions affected by foreground stars and the ring light, which
mainly perturbs the outer regions of the central galaxy along its
major axis, are accurately masked, before the fit is performed. The
structural parameters for bulge and disk components are listed in
Tab.\ref{eso235_2dparam}.  The 2D model, for the central galaxy, in
the J and H bands, and in the B, V, I bands, is a scaled version of
the Kn band model, based on the average colors derived for this
component (given in Tab.\ref{eso235_nir}).  The right panel of
Fig.\ref{eso235_fit} shows the ratio between the whole image for
ESO~235-G58 (i.e. including the outer ring) and the 2D model for the
central galaxy, in the H band. The faint structures related to the
ring are evident in this ``residual image'': these features are found
around the central galaxy and they are elongated toward the NW and SE
directions, at the edge of this component, along the major axis. In
the regions corresponding to such features, in the Kn band, the galaxy
is about $0.8$ mag brighter than the model.\\ The comparison between
the observed and calculated light profiles is shown in the left panel
of Fig.\ref{eso235_fit}: the light profiles along the major axis
deviate from the exponential decrease in the range $10'' \le R \le
20''$, where the galaxy is about $0.6$ to $0.8$ magnitude brighter
than the model.  The analysis performed on the P.A., ellipticity and
surface brightness profiles, described in Sec.\ref{eso235_phot}, lead
us to conclude that the ring light adds on to the central component in
this region. The excess of light, measured on the residual image, is
comparable to those observed along the major axis light profiles
between $10''$ and $20''$.
 
Along the minor axis, in the SW direction, the light distribution 
derived by the fit is brighter than the galaxy light ($\Delta \mu \le 
0.4$ mag), whereas, along the NE directions the galaxy light is more 
luminous than the model ($\Delta \mu \le 0.4$ mag). These differences 
confirm the asymmetry of the minor axis profile, probably due to 
twisted isophotes in the inner regions ($R < 5''$), as already 
mentioned in Sec.\ref{eso235_morph} and Sec.\ref{eso235_dust}. 
 
\begin{figure*}   
\resizebox{\hsize}{!}{   
\includegraphics[width=6cm]{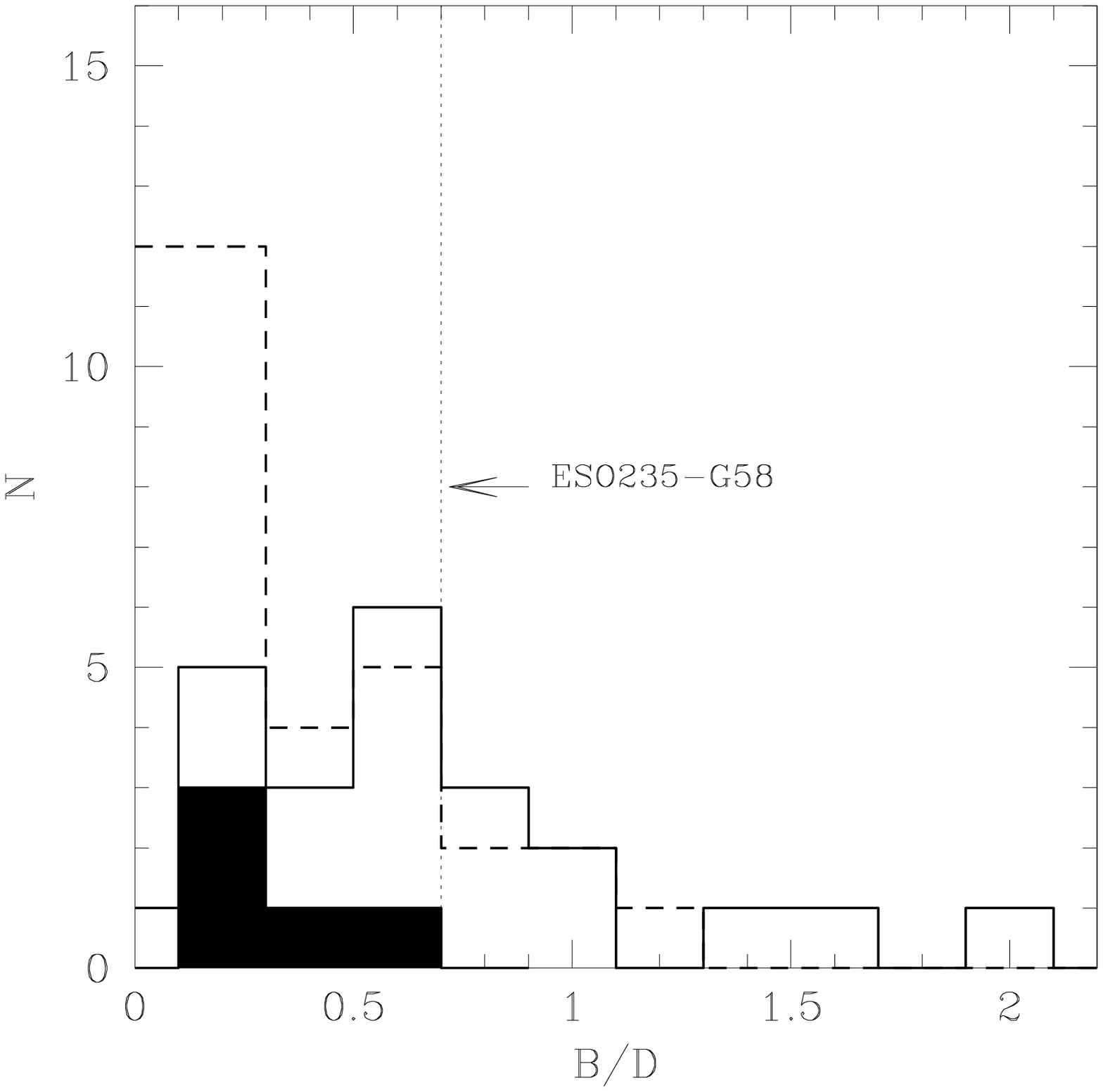} 
\includegraphics[width=6cm]{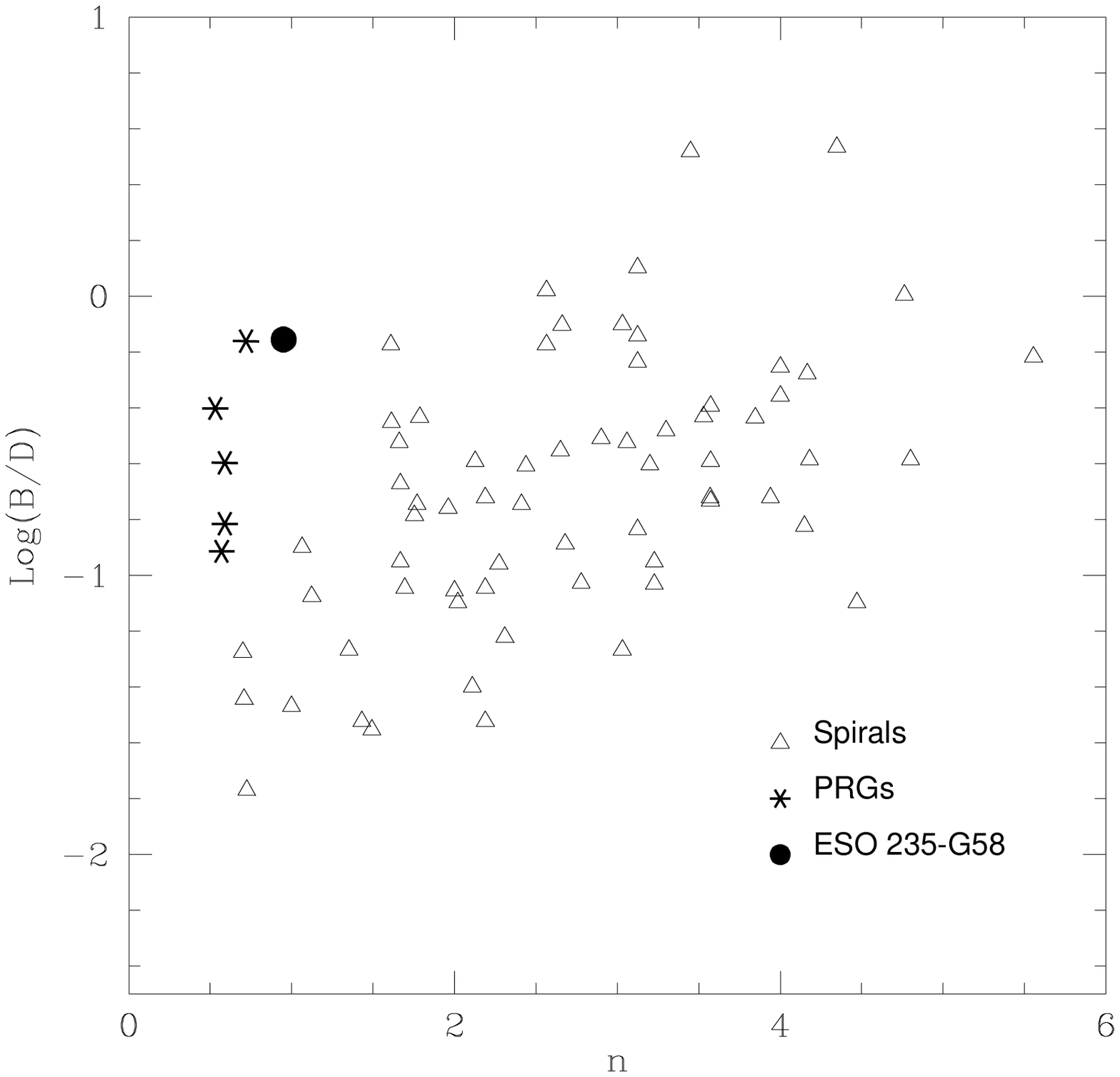} 
}           
\caption{B/D distribution for the PRGs (filled region), by Iodice et   
al. (2002b), for early-type galaxies (continuous line), by Bothun and   
Gregg (1990), and for spiral galaxies (dashed line), by M\"{o}llenhoff   
\& Heidt (\cite{mollenhoff2001}); Khosroshahi et al.   
(\cite{khos2000b}).  The dotted line indicates the value of B/D   
derived for ESO~235-G58 (Tab.\ref{eso235_2dparam}).}   
\label{BD}   
\end{figure*}

By comparing the structural parameters which characterize the light 
distribution of the central galaxy in ESO~235-G58 with the typical 
values for different morphological types of galaxies, including PRGs 
(by Iodice et al. 2002a, 2002b, 2002c), we found that this component 
shows intermediate properties between PRGs and spiral galaxies.  The 
Bulge-to-Disk (B/D) ratio obtained for ESO~235-G58 ($B/D=0.7$), shown 
in Fig.~\ref{BD} (left panel), falls in the range of values typical 
for disk-dominated S0 galaxies and for spiral galaxies; furthermore it 
is also comparable with disk-dominated host galaxies in PRGs, such as 
ESO~415-G26 (see Iodice et al. 2002b). The right panel of 
Fig.~\ref{BD} shows the correlation between the B/D ratio and the $n$ 
exponent. The host galaxy of ESO~235-G58 is characterized by a higher 
value of the B/D ratio with respect to spiral galaxies, for the same 
value of $n$, as it is observed on average for PRGs. This is an effect 
caused by the differences in the disk scalelengths: the disks in 
ESO~235-G58 and in the PRG host galaxies ($<r_{h}>=0.9 \pm 0.5$ kpc) 
are smaller than disks observed in spiral galaxies ($<r_{h}>=4.5 \pm 
2$ kpc, by M\"{o}llenhoff \& Heidt \cite{mollenhoff2001}). 
   
\begin{table}   
\centering   
\caption[]{Structural parameters for the central galaxy, in the Kn band,    
for ESO~235-G58.    
The effective surface brightness $\mu_{e}$ and the central   
surface brightness $\mu_{0}$ are in mag arcsec$^{-2}$, and $\mu_{0}^{c}$   
is corrected for the inclination. $r_{e}$ and $r_{h}$ are respectively   
the effective radius and disk scalelength derived  in arcsec,    
the corresponding values expressed in kpc are derived by using    
$H_{0}=70$ km ${s}^{-1}$ ${Mpc}^{-1}$.}   
\label{eso235_2dparam}   
\begin{tabular}{cc}   
\hline\hline   
Parameter & value\\   
\hline   
$\mu_{e}$ &$15.52\pm 0.08$\\   
$r_{e}$ (arcsec) &$1.84\pm 0.08$\\   
$r_{e}$ (kpc) &$0.54\pm 0.02$\\   
$\mu_{0}$ &$15.00\pm 0.03$\\    
$\mu_{0}^{c}$ &$16.56\pm 0.05$\\   
$r_{h}$ (arcsec) &$4.71\pm 0.06$\\   
$r_{h}$ (kpc) &$1.39\pm 0.02$\\   
$q_{b}$ &$0.88\pm 0.02$\\   
$q_{d}$ &$0.237\pm 0.004$ \\   
$n$ &$0.95\pm 0.08$\\   
$B/D$ &$0.7\pm 0.2$\\   
$\tilde{\chi ^2}$ & 1.1\\   
\hline   
\end{tabular}   
\end{table}              
   
\section{Study of ring light distribution in the B band}\label{eso235_r}   
   
As pointed out in Sec.\ref{eso235_morph}, the apparent inner 
pseudo-ring in ESO~235-G58, which surrounds the central galaxy, is very 
bright in the B and V bands, but it becomes quite faint in the NIR 
bands. Furthermore, it was also stressed that from the ring edges 
emerge two asymmetric spiral arms (Fig.\ref{eso235B}). Thus, we 
studied the ring light distribution in the B band, where this 
component is brighter. The folded light profile in the B band 
was derived from surface brightness profiles 
extracted along the ring major axis in the two main directions 
(Sec.\ref{eso235_phot}, Fig.\ref{eso235_pr}).  By fitting the ring 
light distribution with an exponential law, we derived an estimate of 
the central surface brightness, $\mu_0 = 25.80 
\pm 0.05$ $mag/arcsec^2$, and of the scalelength, $r_h = 23.0 \pm 0.5$   
arcsec; $r_h=6.7 \pm 0.2$ kpc. The scalelength of the ring structure   
in ESO~235-G58 is, on average, larger than the typical values for LSB   
galaxies, which are characterized by a mean value $r_h \sim 2 \pm 1$ kpc,   
in the range $0.15 \le r_h (kpc) \le 9.26$ (by O'Neil et al. 1997);   
whereas it is comparable with the average value observed for spiral   
galaxies, $r_h \sim 6 \pm 3$ kpc ($0.83 \le r_h (kpc) \le 23.5$, by de   
Jong 1996). This component is less extended than the polar structure   
in NGC~4650A, which has a typical scalelength of $8$ kpc in the B band   
(Iodice et al. 2002a).   
   
The ring surface brightness distribution $\mu(r)$ was used to derive   
the $\Delta R /\bar{R}$ ratio, where    
\begin{equation}\label{eq_drr}   
(\Delta R)^2 = \frac{\int_{r_{min}}^{\infty}(r-\bar{R})^{2} \mu(r) dr}   
{\int_{r_{min}}^{\infty} \mu(r) dr}   
\end{equation}   
and $\bar{R}$ is the average radius, weighted by the surface brightness   
distribution, given by   
\begin{equation}   
\bar{R} = \frac{\int_{r_{min}}^{\infty} r \mu(r) dr}{\int_{r_{min}}^{\infty}   
\mu(r)dr}   
\end{equation}   
   
\noindent   
and $r_{min}$ is equal to 3 times the effective radius of the central   
component, in order to exclude the contribution from the    
central galaxy light.   
$\Delta R / \bar{R}$ is a key parameter in studies of polar   
ring stability (see Iodice et al. 2002a, 2002c).    
   
For the ring component in ESO~235-G58, $\Delta R /\bar{R} = 45\%$. 
This value is close to the lower limit in the range of values derived 
for a sample of spiral galaxies (de Jong, 1996), which is $45\% \le 
\Delta R /\bar{R} \le 75\%$. On the other hand, it is near to the   
upper limit in the range of values derived for PRGs, which is $16\%   
\le \Delta R /   
\bar{R} \le 50\%$ (see Iodice et al. 2002a, 2002c): it is comparable   
to the $\Delta R /\bar{R}$ ratio derived for the wide polar ring   
galaxies A0136-0801 ($\sim 45\%$) and NGC~4650A ($\sim 50\%$).   
   
   
\section{Discussion and conclusions}\label{eso235_sum}   
   
We have discussed the NIR and optical properties of the peculiar   
galaxy ESO~235-G58: an accurate photometric study in the optical bands   
led Buta \& Crocker (1993) to classify ESO~235-G58 as an interacting   
system related to the polar ring galaxies. For this peculiar object,   
we have analyzed the light and color distribution in the NIR and   
optical bands and we have compared them with the typical properties   
observed for other morphological galaxy types, including polar ring   
galaxies (described in Iodice et al. 2002a, 2002b, 2002c). The main   
results of this analysis are the following:   
\begin{enumerate}   
\item the P.A. and ellipticity profiles are quite different from those    
observed for nearly face-on barred galaxies, but they are, on the   
other hand, very similar to the typical P.A. and ellipticity profiles   
observed for edge-on disk galaxies;   
\item the high-frequency residual images confirm that the    
central galaxy in ESO~235-G58 has an edge-on disk;   
\item the analysis of the color and light distribution in the central    
component strongly suggests that ESO~235-G58 is not a nearly face-on
barred galaxy and that it shows many similarities to edge-on spiral galaxies
and to the host galaxy in PRGs;
\item the outer ring structure is almost undetectable in the NIR bands: 
it is characterized by very blue colors which are similar to those of
dwarf irregular galaxies and polar rings;
\item the central galaxy in ESO~235-G58 last formed a significant number 
of stars between 1 to 3 Gyrs ago, very similar both to the host galaxy
in PRGs and for spiral galaxies. The last episode of stellar burst in
the ring structure may be as recent as $10^8 yr$, since its colors are
comparable to those of the polar component in NGC~4650A and
ESO~603-G21.
\end{enumerate}   

Furthermore, this peculiar galaxy is characterized by a high amount of
neutral hydrogen, mainly associated with the ring component, as is
also usually observed in many PRGs (van Gorkom et al. 1987; Arnaboldi
et al. 1997; van Driel et al. 2000; 2002): the total HI mass is about
$3 \times 10^9 M_{\odot}$ (Buta \& Crocker 1993; van Driel et
al. 2000; 2002). We derive a total mass for the stellar component of
about $4 \times 10^7 M_{\odot}$, assuming a mass-to-light ratio
$M/L=0.04$ (see Sec.\ref{age_eso235}).  The total mass for the stellar
component of the central galaxy is about $2.1 \times 10^9 M_{\odot}$,
assuming $M/L=1.3$ (see Sec.\ref{age_eso235}).  The total {\it
baryonic mass} in the ring structure, i.e. stellar plus gaseous
component, is then about 1.4 times larger than the total mass of the
stellar component in the central galaxy. A value larger than unity for
the {\it total baryonic mass}-to-{\it stellar mass} ratio is commonly
observed for PRGs (see Iodice et al. 2002a, 2002c).

The low inclination angle between the central galaxy and ring
($40^\circ$, by Buta \& Crocker 1993), which is unusual in PRGs, may
suggest that the interaction is recent and the ring may not yet have
reached a stable configuration or it may be a transient
structure. Connecting ESO235-G58 to PRGs is important to know this.
The ring's low inclination with respect to the equatorial plane of the
central galaxy makes ESO~235-G58 very similar to another {\it
polar-ring-related} object, {\it NGC 660} (van Driel et
al. 1995). According to the latest N-body simulations by Bournaud \&
Combes (2003), such objects are likely to be formed through an
accretion mechanism, where gas is accreted by the host galaxy from a
gas-rich donor. Bournaud \& Combes' scenario can account for both
nearly polar structures, and more shallowly-inclined rings, whose
radial extension $\Delta R /\bar{R}$ can be up to $55\%$.  These
predictions are consistent with the observed values for these
parameters in ESO~235-G58 (see Sec.\ref{intro} and
Sec.\ref{eso235_r}).
  
The stability of such inclined structures depends on the ring mass and
the dark matter distribution; for a ring at low inclination with
respect to the equatorial plane whose mass is 30\% of the visible
total mass in the system, those numerical simulations (Bournaud \&
Combes 2003, see Section 6.3) predict that the ring is unstable, but
will not disrupt within a timescale of 2.5 Gyrs in a nearly spherical
dark halo. In the case of ESO~235-G58, the ring mass (stars plus gas)
is about $60\%$ of the visible total mass in the system, and
information on the dark halo shape can be derived from the position of
ESO~235-G58 in the $L_B - Log(W_{20})$ plane (where $L_B$ is the total
luminosity of the whole system), as described by Iodice et al. (2003).
The $L_B,\,Log(W_{20})$ values for ESO~235-G58 fall on the average
Tully-Fisher relation for bright disk galaxies (see Iodice et
al. 2003, Fig.~2): this implies that the dark halo in ESO~235-G58 may
be nearly spherical.  Thus the ring will probably persist for still
longer than 2.5 Gyrs.

The present work leads us to conclude that ESO~235-G58 is a {\it
polar-ring-related} galaxy, characterized by a low inclined ring/disk
structure. Such inclined structures are rarely observed, because they
are less stable than those nearly polar and they can be easily
confused with barred galaxies, as it seems to have happened in the
case of ESO~235-G58. This peculiar galaxy provides us with an
intriguing face-on view of the low-inclined disk component of a
PRG-related object. Kinematic observations are urgently needed to
further understand the structure of this enigmatic object. The
analysis of high resolution observations would provide further
insights into the effects of the interaction on both the inner
component and the inclined disk.
   
\begin{acknowledgements}  
The authors would like to thank the anonymous referee whose comments 
and suggestions let us improve the presentation of this work. 
E.I. would like to thank G. De Lucia for the help in the use of 
GISSEL. The authors wish to 
thank the Mt. Stromlo and Siding Spring Observatories for the 
observing time at the 2.3m telescope allocated to this project. RB 
acknowledges the support of NSF grant AST-0205143 to the University of 
Alabama and LSS acknowledges support from AST-0139563. 
   
\end{acknowledgements}

\end{document}